\documentclass[aps,prd,preprint,preprintnumbers,superscriptaddress,nofootinbib,longbibliography]{revtex4-1}

\usepackage{graphicx}
\usepackage{color}
\usepackage{physics}
\usepackage[hyperfootnotes=true]{hyperref}

\begin{document}
\title{Probing Quasi-Dirac Neutrino Oscillations at Long Baseline Experiments}
\author{Noor-In\`es Boudjema}
\email{noor-ines.boudjema.19@ucl.ac.uk}
\affiliation{Department of Physics \& Astronomy, University College London, Gower Street, London WC1E 6BT, United Kingdom}

\author{Frank F. Deppisch}
\email{f.deppisch@ucl.ac.uk}
\affiliation{Department of Physics \& Astronomy, University College London, Gower Street, London WC1E 6BT, United Kingdom}

\author{Suka Sriyansu Pattanaik}
\email{suka.sriyansu@gmail.com}
\affiliation{Department of Physics and Astronomy, National Institute of Technology Rourkela, Rourkela 769008, India}

\begin{abstract}
The Dirac or Majorana nature of neutrinos remains one of the most fundamental open problems in particle physics. A natural intermediate scenario arises when small lepton-number-violating Majorana mass terms break the exact Dirac symmetry. These cause mass eigenstates to form nearly degenerate pairs, each separated by a small mass splitting. We study this quasi-Dirac scenario within a five-neutrino framework, extending the Standard Model by two right-handed neutrinos, which form a nearly degenerate sterile pair alongside the three active states. This extended mixing structure introduces additional phenomenological parameters, including new mixing angles, mass splittings, and CP phases. We use appearance and disappearance data from NO$\nu$A and T2K to constrain the active-sterile mixing angles $\theta_{sa}$ and the mass splitting within the sterile pair $\Delta m_{54}^2$ and forecast the sensitivity achievable at DUNE. We analyse how these parameters modify CP-violating observables relative to the standard three-neutrino case, comparing predicted event rates across both mass orderings and as a function of $\delta_{CP}$.
\end{abstract}
\maketitle

\section{Introduction}
The three-neutrino paradigm has proven remarkably successful. Oscillation experiments spanning solar, atmospheric, reactor, and long-baseline accelerators have now determined five of the six parameters of the Pontecorvo–Maki–Nakagawa–Sakata (PMNS) mixing matrix with considerable precision \cite{Fukuda, Kajita:2010zz}. In addition, the discovery of flavour oscillations itself established that neutrinos carry non-zero masses, providing direct evidence for physics beyond the Standard Model (SM). Yet the framework remains incomplete in fundamental respects. The absolute mass scale is unknown, the ordering of the mass eigenstates is only weakly constrained, and the Dirac or Majorana nature of the neutrino is entirely undetermined by oscillation measurements. Central to this open question is the Dirac assumption that neutrinos and antineutrinos are distinct particles of the same mass. If neutrinos are Majorana fermions, this assumption breaks down, and a small lepton-number-violating Majorana mass for the right-handed component $\nu_R$ induces a splitting between two nearly degenerate mass eigenstates. Since $\nu_R$ is a SM gauge singlet, its Majorana mass $m_R$ is technically unconstrained. It is typically considered in a Seesaw scenario where $m_R$ is much larger than the active neutrino masses resulting in a rich phenomenology \cite{Deppisch:2015qwa, Abdullahi:2022jlv}. On the other hand, it can be naturally smaller than the Dirac mass, yielding the quasi-Dirac scenario \cite{Carloni:2025dhv, Dev_2025,Fong_2025}. In this case, two approximate Dirac pairs with a tiny intra-doublet splitting $\delta m^2$ drive distinctive interference effects in oscillation probabilities, and their lepton-number-violating character directly suppresses neutrinoless double beta decay ($0\nu\beta\beta$) relative to the standard Majorana case \cite{valle1983, Gu_2012, Anamiati:2017rxw, Bolton:2019pcu, Anamiati:2019maf, Bolton:2019bou, Bolton:2022tds}. Other lepton-number-violating effects, such as the corresponding oscillations \cite{Bolton:2019wta} are likewise suppressed.

A concrete and theoretically motivated realisation arises in the inverse seesaw (ISS) mechanism \cite{mohapatra1986, Dev:2009aw}, which extends the SM by pairs of sterile fermions carrying a small lepton-number-violating Majorana mass $\mu$ whose smallness is technically natural. In this framework the sterile sector organizes into quasi-Dirac pairs with splitting set by $\mu$, while the low-energy spectrum is captured by an effective $3+2$ neutrino mixing framework. The two additional mass eigenstates, labelled $\nu_4$ and $\nu_5$, are heavier than the three active states and form a quasi-Dirac pair: they are nearly degenerate, with $\Delta m^2_{54} \propto \mu$ controlling a characteristic interference term in oscillation probabilities that has no analogue in standard $3+1$ or $3+2$ scenarios with non-degenerate sterile masses. This interference structure makes long-baseline oscillation experiments a particularly sharp probe \cite{deGouvea:2022kma}. The sensitivity of DUNE to sterile quasi-Dirac neutrinos was investigated by Abada et al.~\cite{Abada:2025edq}, demonstrating that its long baseline of 1285 km and broad energy coverage make it particularly well-suited to constraining the intra-doublet mass splitting, enhancing sensitivity to sub-leading oscillatory effects. Complementary constraints arise from existing data from NO$\nu$A \cite{Abe_2021} and T2K \cite{NOvA:2019cyt}, which probe distinct $L/E$ ranges. In this work we simulate all three experiments using GLoBES, analysing both $\nu_\mu$ disappearance and $\nu_e$ appearance channels, to derive projected sensitivities to the quasi-Dirac mixing parameters and the CP-violating phases they introduce. Here, the near-cancellation between quasi-Dirac pair components generically suppresses the light-neutrino exchange amplitude and makes the residual rate sensitive to the intra-pair splitting, representing a qualitatively different phenomenology from the standard Majorana case.

The paper is organized as follows. Section~\ref{sec::theory} presents the theoretical framework and derives the effective $5\nu$ mixing matrix in the quasi-Dirac limit. Section~\ref{sec::simulation} describes the GLoBES simulations and statistical methodology. Section~\ref{sec::results} presents constraints on the quasi-Dirac parameters, projected sensitivities at DUNE, and a comparative analysis of three-neutrino versus five-neutrino event rate predictions.

\section{Quasi-Dirac Neutrino Oscillations} 
\label{sec::theory}

\subsection{Quasi-Dirac Neutrinos}
While neutrino-antineutrino oscillations are in principle possible for Majorana neutrinos, they are suppressed by $(m/E)^2$ and are therefore unobservable at oscillation experiments \cite{Pontecorvo:1967fh}. The question of the Dirac or Majorana nature of neutrinos therefore remains an open question. Resolving this requires searching for lepton number-violating processes, the most experimentally promising of which is neutrinoless double beta $(0\nu\beta\beta)$ decay. While the Dirac and Majorana cases are often treated as distinct possibilities, the former is more precisely understood as a special limit of the Majorana framework in which all LNV mass terms vanish. In the minimal scenario comprising a single active flavor $\nu_L$ and a right-handed singlet $\nu_R$, the most general neutrino mass Lagrangian in the $(\nu_L,\nu_R^C)^T$ basis is,
\begin{equation}
    \mathcal{L}_{\rm mass} = 
    -\frac{1}{2}\begin{pmatrix} \nu_L & \nu_R^C \end{pmatrix} 
    \underbrace{\begin{pmatrix} \mu_L & m_D \\ m_D & \mu_R \end{pmatrix}}_{\displaystyle\mathcal{M}} 
    \begin{pmatrix} \nu_L \\ \nu_R^C \end{pmatrix} + \text{h.c.},
\end{equation}
where the Majorana masses $\mu_L$ and $\mu_R$ violate lepton number by two units, and the Dirac mass $m_D$ is a Dirac mass generated through the Higgs mechanism. When $\mu_L = \mu_R = 0$, lepton number is conserved and the two mass eigenstates are exactly degenerate. Small nonzero values of $\mu_L$ and $\mu_R$ define the quasi-Dirac scenario~\cite{WOLFENSTEIN1981147, Petcov:2013poa, valle1983}. We factor out $m_D$ and define
\begin{equation} 
    r_\pm \equiv \frac{\mu_R \pm \mu_L}{2m_D}, 
\end{equation}
so that the quasi-Dirac regime is $r_\pm \ll 1$. The exact eigenvalues of $\mathcal{M}$ are 
\begin{equation} 
    m_{1,2} = 
    \frac{\mu_L + \mu_R}{2} \mp \sqrt{m_D^2 + \frac{(\mu_R-\mu_L)^2}{4}}, 
\end{equation}  
which in the quasi-Dirac limit expand to 
\begin{equation} 
    m_{1,2} \approx m_D\left(-1 + r_+ \mp \frac{r_-^2}{2}\right), 
\end{equation}  
and the mixing matrix is
\begin{equation} 
    U \approx \frac{1}{\sqrt{2}}
    \begin{pmatrix} 
        -(1 + r_-/2) & 1 - r_-/2 \\
        1 - r_-/2 & 1 + r_-/2 
    \end{pmatrix}.
\end{equation} 
The two parameters $r_+$ and $r_-$ govern distinct physical effects. $r_+$ controls the intra-doublet mass-squared splitting
\begin{equation} 
    \delta m^2 \equiv 
    m_2^2 - m_1^2 \approx 4m_D^2, \quad r_+ = 2m_D(\mu_R + \mu_L), 
\end{equation}  
which enters oscillation probabilities as an additional period absent from the standard three-flavour framework. $r_-$ controls the deviation of the mixing angles from maximality. Defining the small angle  
\begin{equation} 
    \theta \equiv \frac{r_-}{2} = \frac{\mu_R - \mu_L}{4m_D}, 
\end{equation}
the first row of $U$ becomes $\frac{1}{\sqrt{2}}(-(1+\theta),, 1-\theta)$, so that $\theta = 0$ corresponds to exactly maximal mixing and nonzero $\theta$ shifts the oscillation amplitudes. The two effects are not fully independent: $\mu_L = \mu_R$ gives $r_- = 0$, splitting the spectrum at exactly maximal mixing, while $\mu_L = -\mu_R$ gives $r_+ = 0$ and the leading splitting vanishes, but a residual $\delta m^2 \approx (\mu_R - \mu_L)^2/m_D = 4m_D r_-^2$ survives at second order, suppressed by one further power of $r_-$. In the model presented below we set $\mu_L = 0$. At leading order, allowing both $\mu_L$ and $\mu_R$ to be nonzero would decouple the mass splitting from the mixing angle deviation: since $\delta m^2 \propto \mu_R+\mu_L$ and $\theta \propto\mu_R-\mu_L$, a partial cancellation between the two could suppress one observable while leaving the other unaffected, obscuring the direct link between the lepton-number-violating scale and the quasi-Dirac phenomenology. Setting $\mu_L = 0$ ensures that both $\delta m^2$ and $\theta$ are controlled by the single parameter $\mu_R$, preserving a transparent and predictive connection between lepton number violation and the oscillation observables.

\subsection{Theoretical Framework}
We extend the Standard Model by introducing two SM-singlet Weyl fermions
$N_L$ and $N_R$, alongside three right-handed singlets $\nu_{\alpha R}$
($\alpha = e,\mu,\tau$). The active neutrinos are predominantly Dirac
fermions, acquiring masses $m_{D,\alpha\beta} = y_{D,\alpha\beta}\,v/\sqrt{2}$ after electroweak symmetry breaking. We also assign a Majorana mass $\mu$ to $N_R$. With these definitions, the SM Lagrangian is extended to
\begin{equation}
    \mathcal{L} = 
    \mathcal{L}_{\rm SM} 
    - y_{D,\alpha\beta} \bar{L}_\alpha \tilde{H} \nu_{\beta R} 
    - y_{L,\alpha} \overline{N_L} H^\dagger \nu_{\alpha R} 
    - y_{R,\alpha} \bar{L}_\alpha \tilde{H} N_R 
    - m_S \overline{N_L} N_R 
    - \frac{1}{2}\mu \overline{N_R^c} N_R 
    + \text{h.c.},
\end{equation}
where $L_\alpha = (\nu_{\alpha L},\ell_\alpha)^T$ is the SM lepton doublet, $\tilde{H} = i\sigma_2 H^*$ with $H=(H^+,H^0)^T$, and $m_S$ is the Dirac mass of the sterile sector. After electroweak symmetry breaking, $\langle H^0\rangle = v\approx 246$ GeV, the Yukawa couplings give rise to the Dirac mass matrices $\mathcal{M}_D $, $\mathcal{M}_L$, and $\mathcal{M}_R$, of dimensions $3\times3$, $1\times3$, and $3\times1$ respectively and the Lagrangian gives rise to the following neutrino mass matrix in the basis $(\nu_L, N_L,\nu_R^c,N_R^c)^T$,
\begin{equation}
    \mathcal{M} = \begin{pmatrix}
        0 & 0 & \mathcal{M}_D & \mathcal{M}_R \\
        0 & 0 & \mathcal{M}_L^T & m_S\\
        \mathcal{M}_D^T & \mathcal{M}_L & 0 & 0\\
        \mathcal{M}_R^T & m_S & 0 & \mu
    \end{pmatrix}.
\end{equation}
This matrix can be diagonalised by an $8 \times 8$ unitary matrix $U$ such that
\begin{equation}
    U^T \cdot \mathcal{M}\cdot U = \textnormal{diag}(m_1,...,m_8),
\end{equation}
which yields eight mass eigenvalues. In the limit $\mu=0$, lepton number is conserved and the neutrino mass matrix takes the off-diagonal block form
\begin{align}
    \mathcal{M} = \begin{pmatrix}
        0 & \mathcal{M}_A \\
        \mathcal{M}_A^T & 0
    \end{pmatrix}, \qquad \text{where} \qquad 
    \mathcal{M}_A = \begin{pmatrix}
        \mathcal{M}_D & \mathcal{M}_R \\
        \mathcal{M}_L^T & m_S
    \end{pmatrix},
\end{align}
such that $\mathcal{M_A}$ encapsulates both Dirac-like and sterile mass terms, and being generally complex, it can be diagonalised via a bi-unitary transformation \cite{Giunti:2007ry},
\begin{align}
    V^T \cdot \mathcal{M}_A\cdot W= \textnormal{diag}(m_i) = M_A, \qquad
    W^T \cdot \mathcal{M}_A^T\cdot V = \textnormal{diag}(m_i) = M_A,
\end{align}
with $V$ and $W$ unitary. The transpose convention, rather than the Hermitian conjugate, follows directly from the structure of $\mathcal{M}$. The Majorana mass term $\frac{1}{2}\mu,\overline{N_R^c}N_R$ forces $\mathcal{M}$ to be complex symmetric, $\mathcal{M} = \mathcal{M}^T$, and for complex symmetric matrices the correct decomposition preserving real non-negative eigenvalues is the Takagi decomposition \cite{takagi1925, Horn:1985}, $U^T \mathcal{M} U = M$, rather than the unitarily equivalent $U^\dagger \mathcal{M} U$. This is the standard convention in the neutrino mass literature \cite{Bilenky:1987ty, Giunti:2007ry}, and it propagates to the sub-block level: the block diagonalisation of $\mathcal{M}$ naturally induces the conditions $V^T \mathcal{M}_A W = M_A$ and $W^T \mathcal{M}_A^T V = M_A$ on $\mathcal{M}_A$. The diagonal entries $m_i$ $(i = 1,\ldots,8)$ are the singular values of $\mathcal{M}_A$ and correspond to the physical masses of the active and sterile neutrinos.
\begin{align}
\label{block_diag}
    U_0 = \frac{1}{\sqrt{2}} \begin{pmatrix}
        V & 0\\
        0 & W
    \end{pmatrix} \begin{pmatrix}
        I_4 & -I_4\\
        I_4 & I_4
    \end{pmatrix}, \qquad 
    U_0^T\mathcal{M}U_0 = \begin{pmatrix}
        M_A & 0\\
        0 & -M_A
    \end{pmatrix},
\end{align}
where $I_4$ is the identity matrix of dimension 4. The resulting spectrum consists of eigenvalues in $\pm$ pairs, with each pair corresponding to states of opposite lepton number. To assign positive physical masses and encode the relative CP parities, a diagonal Majorana phase matrix,
\begin{equation}
    D = \textnormal{diag}(e^{i\phi_1/2},...,e^{i\phi_4/2},e^{i\phi_5/2},...,e^{i\phi_8/2}),
\end{equation}
is introduced, with the condition
\begin{align}
    \phi_{j+4} = \phi_j +\pi, \qquad j = 1,\dots,4.
\end{align}
This ensures that the negative eigenvalues are rotated to positive physical values, and the relative phase $\pi$ reflects the opposite CP parity of the two states within each $\pm$ pair. The full diagonalisation is then
\begin{align}
    U = U_0 D, \qquad 
    U^T\mathcal{M}U = \begin{pmatrix}
        M_A && 0\\
        0 && M_A
    \end{pmatrix},
\end{align}
where all physical masses are positive. Each $\pm$ pair thus forms a Dirac neutrino, with the particle and antiparticle components distinguished by their CP parity. Although the full bi-unitary diagonalisation formally involves both $V$ and $W$, all neutrino mass eigenstates are linear combinations of both left-handed $(\nu_L,N_L)$ and right-handed $(\nu_R,N_R)$ fields. Consider the charged current,
\begin{equation}
    \mathcal{L}_{CC} = \overline{L}\gamma^\mu\nu_LW_\mu,
\end{equation}
which does not involve right-handed neutrinos. It is therefore possible to proceed through a rephasing of the right-handed fields,
\begin{equation}
\label{eq::rephase}
    \nu_R' = W^\dagger \begin{pmatrix}
        \nu_R \\
        N_R 
    \end{pmatrix}.
\end{equation}
In this case the new mass matrix can be written as
\begin{align}
    \mathcal{M}' = \begin{pmatrix}
        0 & \mathcal{M}_A'\\
        \mathcal{M}_A'^T & 0
    \end{pmatrix}, \qquad 
    \mathcal{M}_A' = \mathcal{M}_AW^\dagger.
\end{align}
In this basis, the diagonalisation from Equation \eqref{eq::rephase} becomes,
\begin{align}
    V^T\mathcal{M}_A'  = \textnormal{diag}(m_i).
\end{align}
Through this rotation, $W$ can be absorbed into the sterile sector without affecting the active components, so the oscillation probabilities are entirely determined by $V$. Since $\mathcal{M}$ is block off-diagonal, lepton number is conserved and the system describes four Dirac neutrinos. A $4\times4$ unitary matrix has 10 phases, of which 7 can be removed by rephasing the left- and right-handed fields, leaving 3 physical Dirac phases. $V$ is therefore described by 6 mixing angles and 3 Dirac phases,
\begin{equation}
    V =  R^{34}W^{24}R^{23}W^{14}W^{13}R^{12}.
\end{equation}
Here, $W^{ab}$ is a complex rotation in the $a$-$b$ plane depending on the mixing angle $\theta_{ab}$ and Dirac phase $\eta_{ab}$, with elements \cite{Giunti:2007ry}
\begin{equation}
    \label{Wab}
    [W^{ab}(\theta_{ab},\eta_{ab})]_{rs} = \delta_{rs}+(c_{ab}-1)(\delta_{ra}\delta_{sa}+\delta_{rb}\delta_{sb})+s_{ab}(e^{i\eta_{ab}}\delta_{ra}\delta_{sb}-e^{-i\eta_{ab}}\delta_{rb}\delta_{sa}),
\end{equation}
where $c_{ab}$ and $s_{ab}$ denote the cosine and sine of $\theta_{ab}$. $R^{ab}$ is equal to $W^{ab}$ with $\eta_{ab} = 0$. The rephasing freedom of the right-handed fields allows three of the six phases in the decomposition to be set to zero. We use this freedom to remove the phases in the $3$-$4$, $2$-$3$, and $1$-$2$ planes, so that $R^{34}$, $R^{23}$, and $R^{12}$ are real rotations, leaving the three physical Dirac phases $\eta_{13}$, $\eta_{14}$, and $\eta_{24}$ carried by $W^{13}$, $W^{14}$, and $W^{24}$.
The block off-diagonal structure and the~$\pm$~eigenvalue pairing rely crucially on exact lepton number conservation. When $\mu \neq 0$, lepton number conservation is broken and $\mathcal{M}$ can no longer be written in block off-diagonal form, so the simple diagonalisation described above does not apply and the $\pm$ structure of eigenvalues is lost. To determine the physical spectrum in this case, it is convenient to re-express $\mathcal{M}$ in the $(\nu_L,\nu_R,N_L,N_R)$ basis,
\begin{align}
    \mathcal{M}= \begin{pmatrix} 0&\mathcal{M}_D & 0 & \mathcal{M}_L\\ \mathcal{M}_D^T & 0 & \mathcal{M}_R&0\\ 0 & \mathcal{M}_R^T & 0& m_S\\ \mathcal{M}_L^T& 0 &m_S&\mu \end{pmatrix},
\end{align}
and partition it into "large", "small", and "interference" blocks $L$, $S$, and $R$. Setting $\mu = 0$,
\begin{align}
    \mathcal{M} = \begin{pmatrix} S & R\\ R^T & L \end{pmatrix}, \quad 
    S = \begin{pmatrix} 0 & \mathcal{M}_D \\ \mathcal{M}_D^T & 0 \end{pmatrix}, \quad 
    L= \begin{pmatrix} 0 & m_S\\ m_S &0 \end{pmatrix}, \quad 
    R = \begin{pmatrix} 0 & \mathcal{M}_L \\ \mathcal{M}_R &0 \end{pmatrix}.
\end{align}
In the limit $m_S \gg 1$, the effective light neutrino mass matrix is obtained via the Schur complement,
\begin{align}
    S_\text{eff} = S - R^T L^{-1} R = \begin{pmatrix} 0 & \mathcal{M}_D'\\ \mathcal{M}_D'^T & 0 \end{pmatrix}, \qquad \mathcal{M}_D' = \mathcal{M}_D - \frac{1}{m_S}\mathcal{M}_L^T\mathcal{M}_R,
\end{align}
where the physical light neutrino masses $m_i$ are the singular values of $\mathcal{M}_D'$, i.e. the positive square roots of the eigenvalues of $\mathcal{M}_D'\mathcal{M}_D'^T$. The eigenvalues of $S_{\text{eff}}$ come in pairs $\pm m_i$, and the heavy sterile masses are simply equal to $\pm m_S$, and as outlined previously, oscillation probabilities can be entirely determined by a $4\times4$ matrix.

When $\mu \neq 0$, it enters as a small perturbation through the modified heavy block inverse,
\begin{align}
    L = \begin{pmatrix} 0 & m_S \\ m_S & \mu \end{pmatrix}, \qquad 
    L^{-1} = \begin{pmatrix} -\mu/m_S^2 & 1/m_S \\ 1/m_S & 0 \end{pmatrix},
\end{align}
which modifies the Schur complement to
\begin{align}
    S_\text{eff}' = \begin{pmatrix} \mu_{\text{eff}} & \mathcal{M}_D'\\ \mathcal{M}_D'^T & 0 \end{pmatrix}, \qquad 
    \mu_{\text{eff}} = \frac{\mu}{m_S^2}\mathcal{M}_R\mathcal{M}_R^T.
\end{align}
The parameter $\mu$ enters solely through $\mu_{\text{eff}}$ in the upper-left block, which constitutes a perturbation to $S_{\text{eff}}$ of order $\mu/m_S^2$. By first-order perturbation theory, the shift in the $i$-th active neutrino mass is
\begin{align}
    \delta m_i = \langle \psi_i |\, \delta S\, | \psi_i \rangle 
    = \frac{\tilde{\mu}_{ii}}{2}, \qquad \tilde{\mu}_{ii} 
    \equiv \left(H^\dagger\, \mu_{\text{eff}}\, H\right)_{ii},
\end{align}
where $H$ diagonalises $\mathcal{M}_D'\mathcal{M}_D'^T$ and the factor of $1/2$ arises from the normalisation of the eigenstates of the block off-diagonal $S_{\text{eff}}$. The perturbed light masses are therefore
\begin{equation}
    m_i' = \pm m_i + \frac{\tilde{\mu}_{ii}}{2}.
\end{equation}
This shift is doubly suppressed by $\mu/m_S \ll 1$ and $\|\mathcal{M}_R\|^2/m_S^2 \ll 1$, so the active spectrum is suppressed at first order in $\mu$. The leading effect of $\mu$ appears instead in the heavy sterile sector. The effective heavy matrix
\begin{align}
    L_\text{eff} = \begin{pmatrix} 0 & m_S\\ m_S & \mu \end{pmatrix}
\end{align}
yields sterile masses $m_{4,5} \approx \pm m_S + \mu/2$, giving a mass splitting $\Delta m_{54} = \mu$. This lifts the degeneracy between the two sterile states into a quasi-Dirac pair with opposite CP parities, with the splitting providing a direct measure of the degree of lepton number violation. The spectrum therefore reduces to five distinct masses: the three light neutrinos and two quasi-Dirac sterile states. In this regime the mixing matrix $V$ becomes a $5\times5$ matrix comprising 10 mixing angles and 6 Dirac phases \cite{Abada:2025edq},
\begin{equation}
    U =  
    W^{45}R^{35}R^{34}W^{25}W^{24}R^{23}W^{15}W^{14}W^{13}R^{12} \cdot D_M,
\end{equation}
where $D_M$ is the diagonal matrix of Majorana phases. The relevant parametrisation now corresponds to three Dirac neutrinos and two Majorana neutrinos, requiring two independent Majorana phases,
\begin{equation}
    D_M = \textnormal{diag}(1,1,1,e^{i\phi_4/2},e^{i\phi_5/2}).
\end{equation}
To ensure that the sterile neutrinos form a quasi-Dirac pair with opposite CP parities, we impose
\begin{equation}
    \phi_5 = \phi_4+\pi.
\end{equation}
\begin{figure}[t!]
    \centering
    \includegraphics[width=0.8\linewidth]{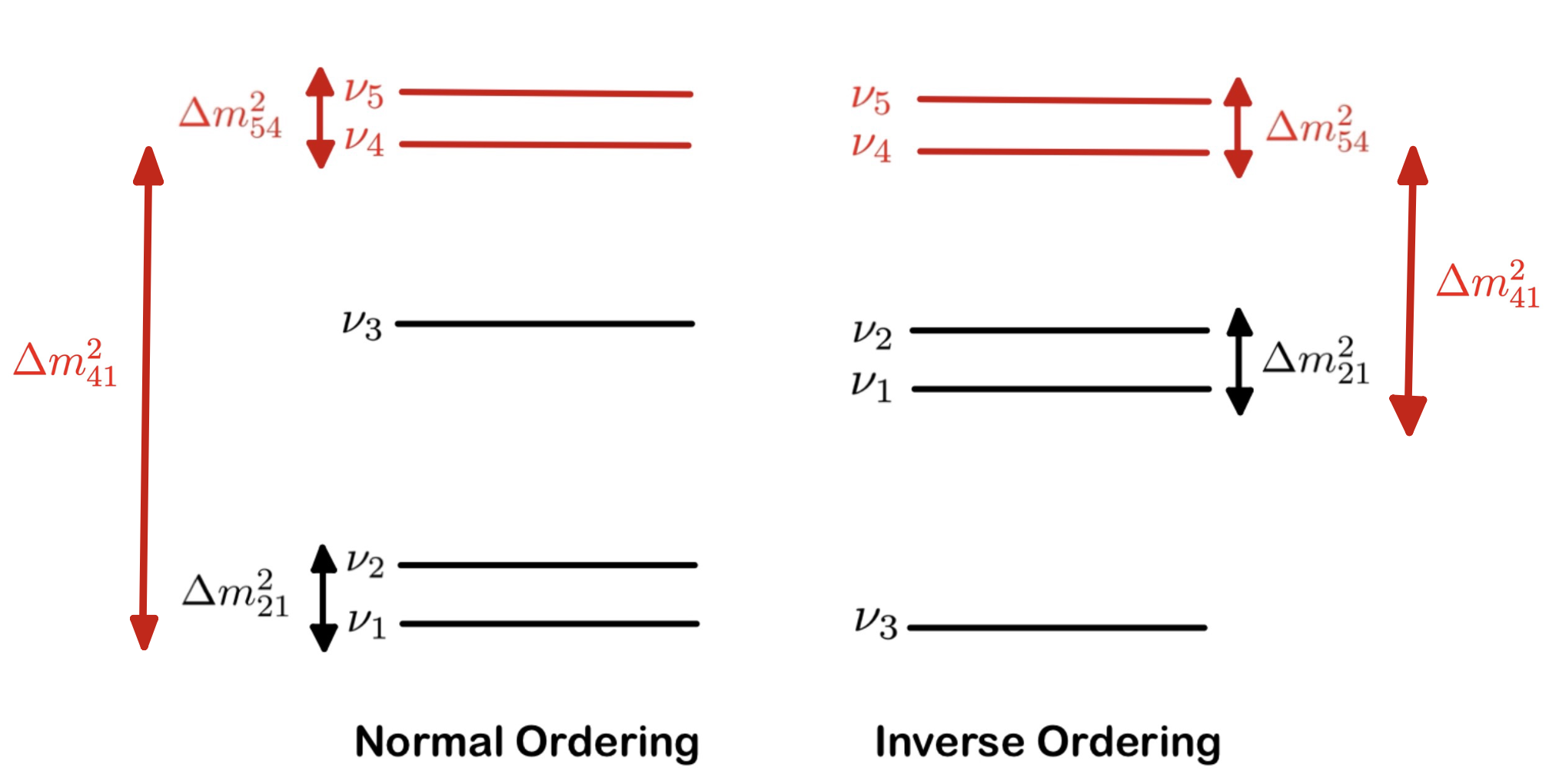}
    \caption{Ordering of the active neutrino masses $\nu_1,\nu_2,\nu_3$ in the presence of the sterile~neutrinos~$\nu_4,\nu_5$. To render the quasi-Dirac scenario for the sterile neutrinos, we choose the sterile mass splitting $\Delta m_{54}^2$ to be of approximately the same range as the small active mass splitting $\Delta m_{21}^2$.}\label{NO_IO}
\end{figure}
With this parametrisation, 15 new parameters are introduced relative to the standard three-flavour Dirac PMNS matrix. When $\Delta m_{54}^2 = 0$, the quasi-Dirac structure breaks down and the model can be described through a $4\times4$ matrix. It is worth noting that in the limit $\theta_{14},\theta_{24},\theta_{34},\theta_{15},\theta_{25},\theta_{35},\theta_{45} \to 0$, the standard $3\nu$ oscillation is recovered and $\nu_{1,2,3}$ become linear superpositions of only the active states $\nu_{e,\mu,\tau}$. We work in the regime where these new mixing angles all take the same value $\theta_{sa}$ and are small relative to the standard angles $\theta_{12},\theta_{13},\theta_{23}$. The active neutrino masses follow either normal (NO) or inverted (IO) ordering, while the sterile states are assumed to be significantly heavier, with $\Delta m_{41}^2 = 10~\textnormal{eV}^2$ and a mass splitting $\Delta m_{54}^2$ comparable in magnitude to $\Delta m_{21}^2$, as illustrated in Fig.~\ref{NO_IO}. 

These mass splittings lie within the sensitivity range of the KATRIN experiment, which searches for a kink-like distortion in the tritium $\beta$-decay spectrum induced by a fourth mass eigenstate. Based on 259 days of data, KATRIN excludes sterile-to-active mixing angles above a few percent across sterile-to-active mass splittings ranging from a fraction of an eV$^2$ to several hundred eV$^2$ \cite{KATRIN:2025lph}. However, KATRIN is exclusively sensitive to the electron-flavour component of the active-sterile mixing, meaning that the operative constraint applies only to the combination $|U_{e4}|^2 + |U_{e5}|^2 \sim \sin^2\theta_{14} + \sin^2\theta_{15}$. The remaining mixing angles $\theta_{24}, \theta_{25}, \theta_{34}$, and $\theta_{35}$ are invisible to KATRIN and may freely take values in the range $\theta_{sa} \in [0.1, 1]$ without tension with this bound. The quasi-Dirac structure of the sterile sector introduces an additional subtlety: since $\nu_4$ and $\nu_5$ are nearly degenerate, KATRIN would not resolve the two states individually, and the bound applies to the incoherent sum $\sin^2\theta_{14} + \sin^2\theta_{15}$ rather than to either angle alone. Consequently, while the electron-flavour mixing angles are required to remain small, the bulk of the six-dimensional active-sterile mixing parameter space parametrised by $\theta_{sa}$ remains viable within current experimental constraints. 

\subsection{Quasi-Dirac Sterile Neutrino Oscillations at Long-baseline Experiments}\label{QD_LBL_Theory}
\begin{figure}[t!]
    \centering
    \includegraphics[width=0.49\linewidth]{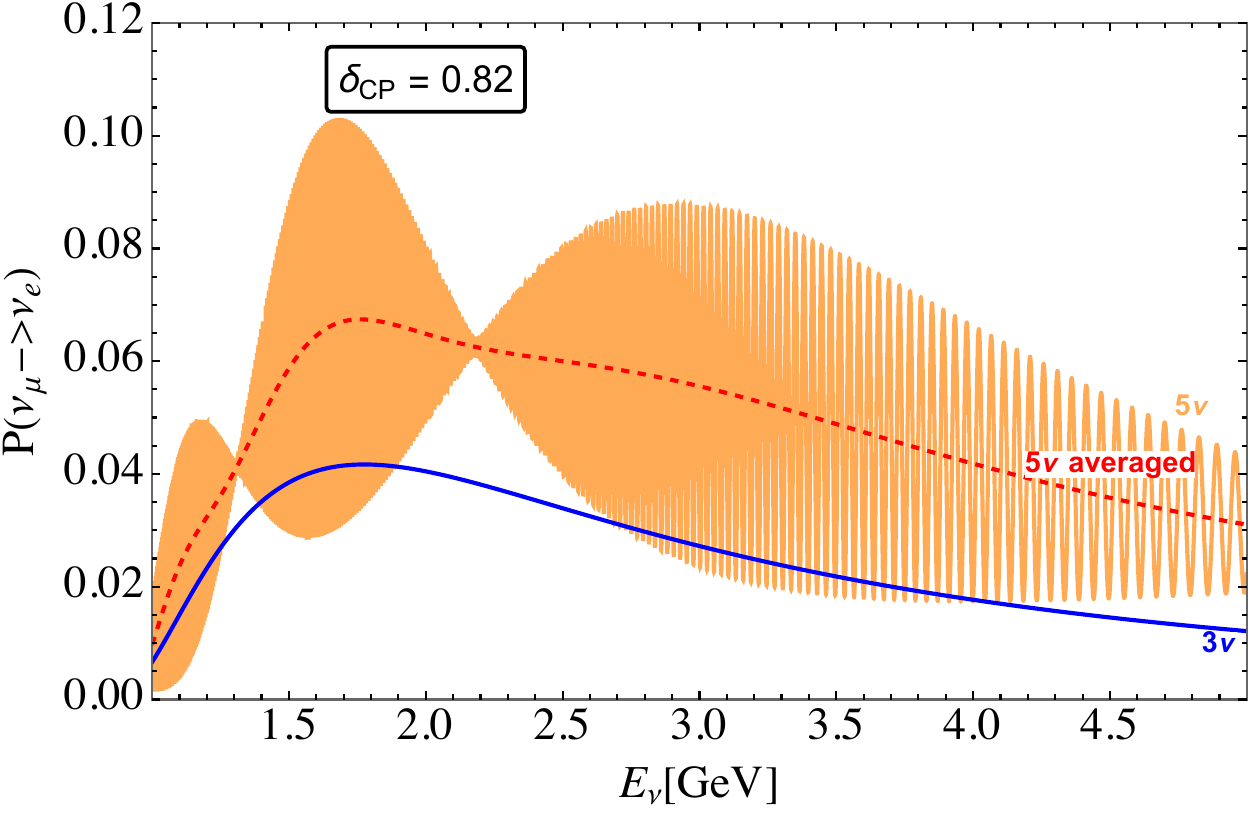}
    \includegraphics[width=0.49\linewidth]{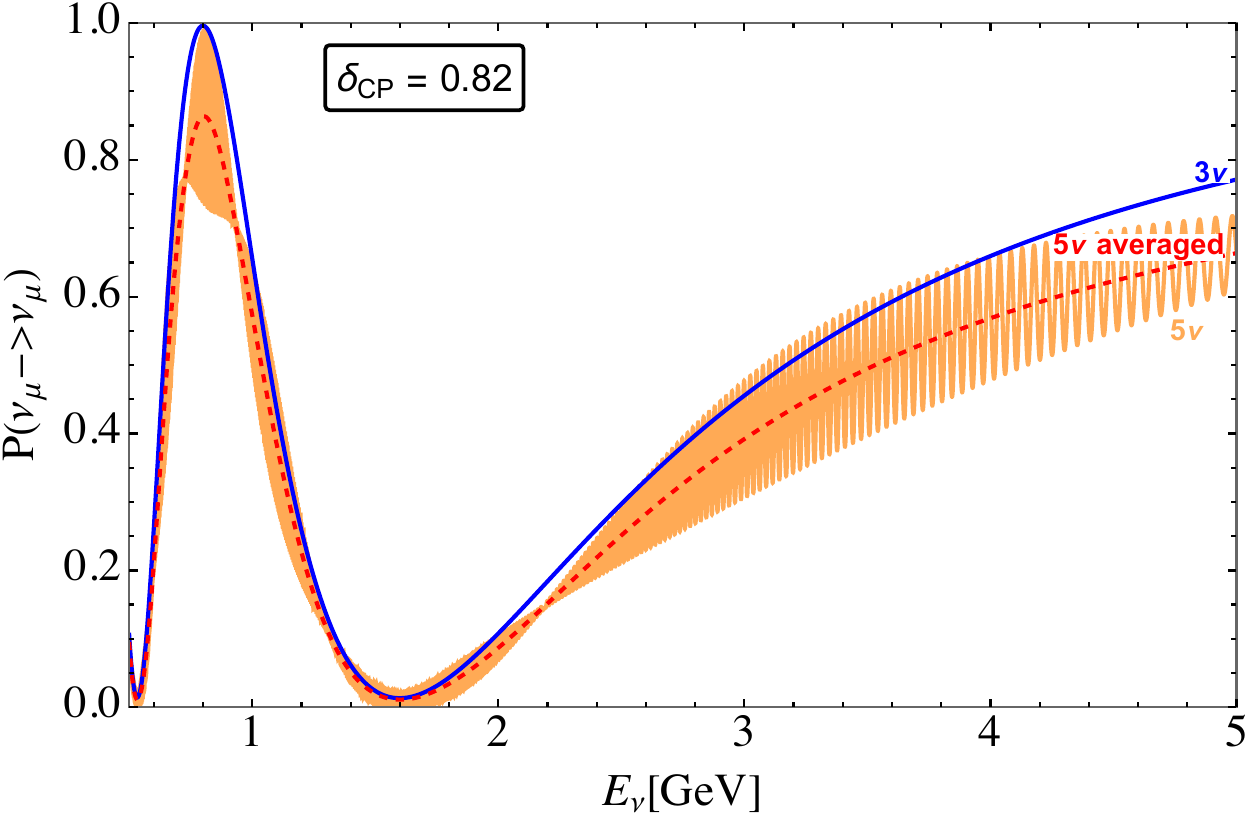}
    \includegraphics[width=0.49\linewidth]{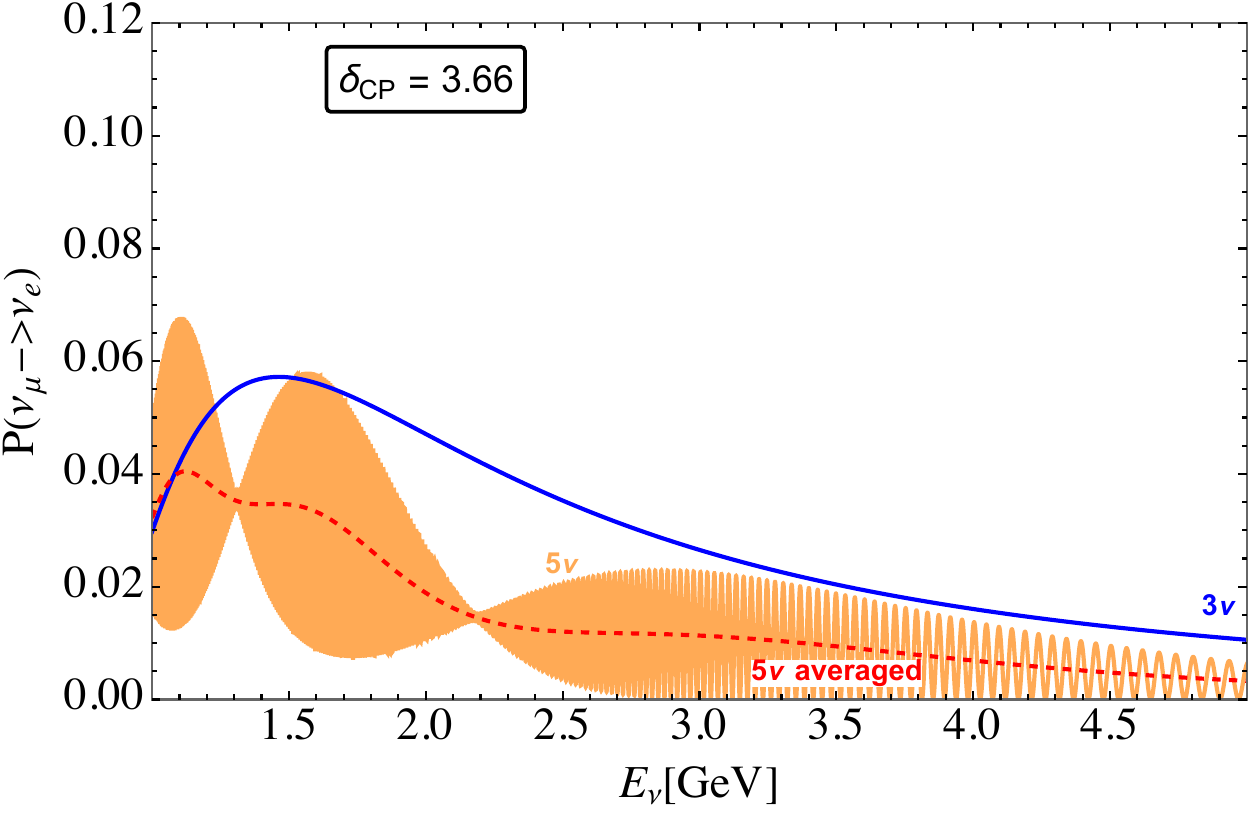}
    \includegraphics[width=0.49\linewidth]{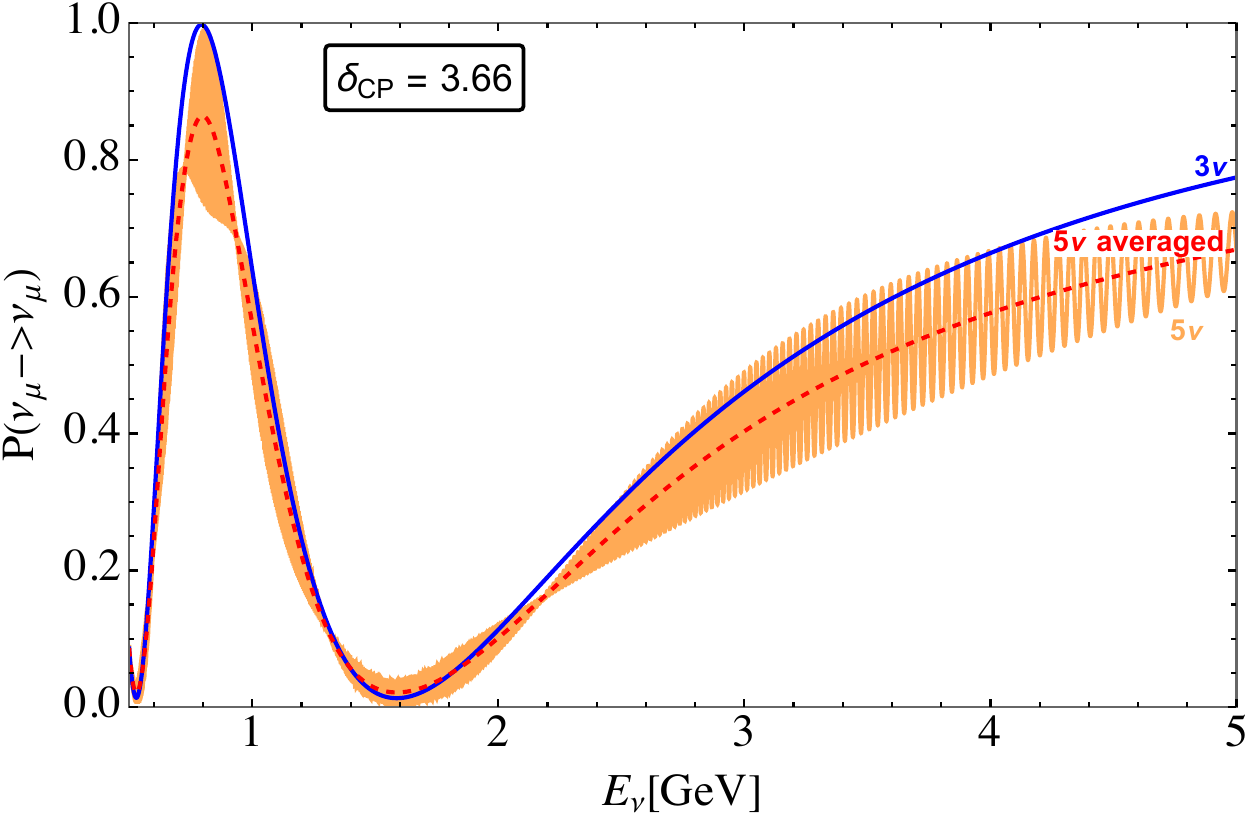}
    \caption{Oscillation probabilities $P(\nu_\mu\to\nu_e)$ (left column) and $P(\nu_\mu\to\nu_\mu)$ (right column) in the SM $3\nu$ and in the $5\nu$ framework over a distance $L = 810$ km, with sterile mass splitting $\Delta m^2_{54}=0.01~\text{eV}^2$, sterile mixing $\theta_{sa}=0.2$, with all other parameters given in Table~\ref{table_params}, for $\delta_{CP} = 0.82$ (top row) and $\delta_{CP} = 3.66$ (bottom row) assuming NO.}
    \label{fig:nova_osc_deltacp_3.66}
\end{figure}

While oscillation experiments cannot distinguish between Dirac and Majorana neutrinos, once we introduce small sources of lepton number violation, oscillation probabilities change \cite{Anamiati:2017rxw}. The probability of a neutrino propagating with energy $E$ from one weak eigenstate $\nu_\alpha$ into $\nu_\beta$ over a distance $L$ in vacuum is,
\begin{align}
\label{oscillation_eqn}
    P(\nu_\alpha \to \nu_\beta) \approx 
    \delta_{\alpha \beta} &- 4\sum_{i>j}^5\Re{U_{\alpha i}U_{\beta i}^*U_{\alpha j}^* U_{\beta j}}\sin^2(x_{ij})\nonumber\\
    &+ 2 \sum_{i>j}^5\Im{U_{\alpha i}U_{\beta i}^*U_{\alpha j}^* U_{\beta j}}\sin(2x_{ij}),
\end{align}
with $x_{ij} = \frac{\Delta m_{ij}^2 L}{4E} = \frac{(m_i^2 - m_j^2) L}{4E}$. As illustrated in Fig.~\ref{NO_IO}, the introduction of two sterile neutrinos gives rise to three distinct classes of mass splittings. The splittings among the three active neutrinos, $\Delta m_{ij}^2$ with $i,j \leq 3$, govern the standard three-neutrino oscillations. The cross-splittings $\Delta m_{i4}^2$ and $\Delta m_{i5}^2$ with $i \leq 3$ are assumed large, with $\Delta m_{41}^2 = 10~\text{eV}^2$ as given in Table~\ref{table_params}, driving rapid high-frequency oscillations visible as the dense orange band in Fig.~\ref{fig:nova_osc_deltacp_3.66}. The corresponding oscillation length,
\begin{align}
    L_\text{osc} = 
    \frac{4\pi E}{\Delta m_{41}^2} 
    \approx 0.5~\text{km} \left(\frac{E}{1~\text{GeV}}\right),
\end{align}
is much shorter than the baseline $L = 810$ km, so these oscillations are unresolvable at long-baseline experiments and average out to a constant contribution, visible as the red line in Fig.~\ref{fig:nova_osc_deltacp_3.66}. The remaining splitting $\Delta m_{54}^2$ characterises the near-degeneracy of the two sterile states, and is taken to be of the same order as the active-sector splittings. The interference between the two sterile states modulates the envelope of the rapid oscillations, shifting the averaged probability away from the pure $3\nu$ prediction in both the appearance channel $P(\nu_\mu \to \nu_e)$ and the survival channel $P(\nu_\mu \to \nu_\mu)$, most prominently near the oscillation maximum at $E_\nu \sim 1.5$ GeV. It is this $\Delta m_{54}^2$-driven shift that constitutes the observable signature of quasi-Dirac neutrinos at long-baseline experiments.

To isolate this effect analytically, we average the full oscillation probability over the rapidly varying phases $x_{41}$, $x_{42}$, and $x_{43}$. Using the unitarity condition $UU^\dagger = I$ and noting that $x_{51} = x_{41} + x_{54}$, the relevant averages are
\begin{align}
    \frac{1}{2\pi} \int_{0}^{2\pi}\sin^2(x_{41}) \, dx_{41} = \frac{1}{2\pi} \int_{0}^{2\pi}\sin^2(x_{51}) \, dx_{41} &= \frac{1}{2}, \nonumber\\
    \frac{1}{2\pi} \int_{0}^{2\pi}\sin(2x_{41}) \, dx_{41} = \frac{1}{2\pi} \int_{0}^{2\pi}\sin(2x_{51}) \, dx_{41} &= 0, 
\end{align}
where the result for $x_{51}$ follows since the constant offset $x_{54}$ does not affect the average over a full period. The $\sin^2$ terms each contribute $\frac{1}{2}$, while all $\sin(2x_{ij})$ interference terms involving the large splittings vanish. The only surviving dependence on the sterile sector then enters through the slowly varying phase $x_{54} = \frac{\Delta m_{54}^2 L}{4E}$, which is of order unity across the relevant energy range and baseline. Neglecting the MSW effect, we define \cite{Karagiorgi:2006jf},
\begin{align}
    \phi_{45} = \arg(U_{\mu5}^* U_{e5}U_{\mu4}U_{e4}^*) 
    \approx (\eta_{14}-\eta_{15}) - (\eta_{24} - \eta_{25}),
\end{align}
where $\eta_{14}, \eta_{24}, \eta_{15}, \eta_{25}$ correspond to the Dirac phases of the sterile neutrinos, such that $\phi_{45} \to 0$ in the limit that the sterile neutrinos form a Dirac pair. The averaged oscillation probability in the $\nu_\mu \to \nu_e$ appearance channel is then
\begin{align}
\label{eq:nue_appearance}
    P(\nu_\mu \to \nu_e) = P(\nu_\mu \to \nu_e)_{3\nu} &+ 2\abs{U_{\mu4}}^2\abs{U_{e4}}^2 + 2\abs{U_{\mu5}}^2\abs{U_{e5}}^2 \nonumber\\
    &+ 4\abs{U_{\mu4}}\abs{U_{e4}}\abs{U_{\mu5}}\abs{U_{e5}} \cos(x_{54}) \cos(x_{54}-\phi_{45}),
\end{align}
where the Dirac phases enter explicitly through $\phi_{45}$. On the other hand, the survival probability $\nu_\mu \to \nu_\mu$ is,
\begin{align}
\label{eq:numu_survival}
    P(\nu_\mu \to \nu_\mu) &= P(\nu_\mu \to \nu_\mu)_{3\nu} + 2\abs{U_{\mu 4}}^2\abs{U_{\mu 5}}^2\cos^2(x_{54}) \nonumber\\
    &\quad - 2\abs{U_{\mu 4}}^2(1-\abs{U_{\mu 4}}^2) - 2\abs{U_{\mu 5}}^2(1-\abs{U_{\mu 5}}^2),\\
    &\approx P(\nu_\mu \to \nu_\mu)_{3\nu} - 2\abs{U_{\mu 4}}^2 - 2\abs{U_{\mu 5}}^2,\nonumber
\end{align}
where the approximation holds when $\abs{U_{\mu4}}^2, \abs{U_{\mu5}}^2 \ll 1$ and the $\cos^2(x_{54})$ term averages to $\frac{1}{2}$. The effect of the sterile mixing in this channel is qualitatively different: the $\nu_\mu$ can now oscillate into the sterile states, reducing the survival probability by an amount proportional to $\abs{U_{\mu4}}^2 + \abs{U_{\mu5}}^2$. This depletion carries no dependence on the Dirac phases, in contrast to the appearance channel, making the two channels complementary probes of the sterile sector. In the quasi-Dirac limit $\abs{U_{\mu4}} = \abs{U_{\mu5}}$ and $\abs{U_{e4}} = \abs{U_{e5}}$, Eqs.~\eqref{eq:nue_appearance} and \eqref{eq:numu_survival} reduce to
\begin{align}
    P(\nu_\mu \to \nu_e) &= P(\nu_\mu \to \nu_e)_{3\nu} + 4\abs{U_{\mu4}}^2\abs{U_{e4}}^2\left(1 + \cos(x_{54})\cos(x_{54}-\phi_{45})\right), \\
    P(\nu_\mu \to \nu_\mu) &= P(\nu_\mu \to \nu_\mu)_{3\nu} - 4\abs{U_{\mu4}}^2.
\end{align}
The sterile contribution here represents a genuine new oscillation channel: $\nu_\mu$ can now oscillate into $\nu_e$ via the sterile states, with an amplitude proportional to the product of the $\nu_\mu$-sterile and $\nu_e$-sterile mixing elements $\abs{U_{\mu4}}\abs{U_{e4}}$ and $\abs{U_{\mu5}}\abs{U_{e5}}$. This is an additive enhancement of the appearance probability, modulated by the interference phase $x_{54}$ and the relative CP phase $\phi_{45}$. We further note that the ratio
\begin{align}
    \frac{4\abs{U_{\mu4}}\abs{U_{e4}}\abs{U_{\mu5}}\abs{U_{e5}}}{2\abs{U_{\mu4}}^2\abs{U_{e4}}^2+2\abs{U_{\mu5}}^2\abs{U_{e5}}^2}
    \approx 2\frac{s_{15}s_{14}s_{25}s_{24}}{s_{15}^2s_{25}^2+s_{14}^2s_{24}^2},
\end{align}
equals unity in the Dirac limit, assuming equal mixing of both sterile neutrinos to each active flavour. The oscillation parameters used throughout this work are given in Table~\ref{table_params}, and the resulting oscillation probabilities for representative values of $\delta_{CP}$ are shown in Fig.~\ref{fig:nova_osc_deltacp_3.66}.

\begin{table}[t!]
\begin{center}
\begin{tabular}{c|c|c}
 Parameter & NO & IO \\ \hline
 $\sin^2{\theta_{12}}$  & 0.307 & 0.308 \\
 $\sin^2{\theta_{23}}$ & 0.57 & 0.47 \\
 $\sin^2{\theta_{13}}$  & 0.022 & 0.022 \\
 $\delta_{CP}$  & 3.66 & 3.66 \\
 $\frac{\Delta m_{21}^2}{10^{-5}~\text{eV}^2}$  & 7.53 & 7.49 \\
 $\frac{\Delta m_{31}^2}{10^{-3}~\text{eV}^2}$ & +2.51 & -2.51  \\ \hline
$\theta_{sa}$& 0.1 - 1 & 0.1 - 1 \\
 $\frac{\Delta m_{41}^2}{\text{eV}^2}$ & $10$ & $10$ \\
 $\frac{\Delta m_{54}^2}{\text{eV}^2}$& $10^{-5}$ - $10^{-1}$ & $10^{-5}$ - $10^{-1}$
\end{tabular}
\end{center}
\caption{Oscillation parameters used in this paper unless indicated otherwise. The columns correspond to the normal (NO) and inverted (IO) mass ordering. The $3\nu$ parameters are taken from \cite{Esteban:2024eli}.}
\label{table_params}
\end{table}
The question of whether $P(\nu_\mu \rightarrow \nu_e)$ is enhanced or suppressed in the 5$\nu$ framework relative to the 3$\nu$ case is not universal, but rather depends sensitively on the value of $\delta_{CP}$. To understand this, one must examine how the CP-violating phase enters the oscillation amplitude in each case. In the standard three-flavour framework, the appearance probability $P(\nu_\mu \rightarrow \nu_e)$ receives contributions from terms that depend explicitly on $\delta_{CP}$, whose sign and magnitude determine whether CP violation acts to enhance or suppress the transition. In the 5$\nu$ framework, these same $\delta_{CP}$-dependent contributions are modified by the presence of the sterile-active mixing. Specifically, the terms in the oscillation amplitude that carry dependence on the CP phase can be collected and written as 
\begin{equation}
    -c_{sa} s_{13}^2 s_{23}^2 c_{13} e^{-i\delta_{CP}} 
    \left(  c_{12}^2 + e^{-ix_{12}} s_{12}^2 - e^{-ix_{13}} \right), 
\end{equation}
where $c_{sa}$ is the cosine of the sterile-active mixing angle. This expression encodes the total contribution to the transition amplitude from all interference terms that depend on $\delta_{CP}$. The overall factor of $c_{sa}$ is the crucial feature: since $c_{sa} < 1$ for any non-zero sterile-active mixing, the magnitude of the CP-sensitive part of the amplitude is universally suppressed in the 5$\nu$ case relative to the 3$\nu$ case, regardless of the value of $\delta_{CP}$ itself. Whether this suppression results in a larger or smaller appearance probability depends on the sign of the real part of the full CP-dependent term when added to the transition probability. Defining the bracketed kinematic expression as
\begin{equation} 
\label{eq:CP_Pro_Expression}
    \mathcal{B} =  c_{12}^2 + e^{-ix_{12}} s_{12}^2 - e^{-ix_{13}}, 
\end{equation} 
this is a complex number whose argument $\arg(\mathcal{B})$ is fixed entirely by the mass eigenvalues, the baseline, and the neutrino energy. The sign of the real part of $-e^{-i\delta_{CP}}\mathcal{B}$, which determines whether the CP-dependent terms are constructive or destructive, changes when $\delta_{CP} + \arg(\mathcal{B}) = \pi$. This defines a threshold,
\begin{equation} 
    \delta_{CP}^{\rm thresh} = \pi - \arg(\mathcal{B}), 
\end{equation} 
which depends on the experimental baseline and neutrino energy through $\arg(\mathcal{B})$. For $\delta_{CP} < \delta_{CP}^{\rm thresh}$, the real part of $-e^{-i\delta_{CP}}\mathcal{B}$ is negative, meaning the CP-dependent terms destructively interfere with the remaining contributions and suppress the appearance probability. Their suppression by $c_{sa}$ in the 5$\nu$ framework therefore partially quenches this destructive interference, raising the transition probability above the three-flavour prediction, as observed for $\delta_{CP} = 0.82$. For $\delta_{CP} > \delta_{CP}^{\rm thresh}$, the real part becomes positive, meaning the CP-dependent terms constructively enhance the appearance probability. In the 5$\nu$ framework this constructive enhancement is suppressed by $c_{sa}$, and the transition probability is correspondingly smaller than the three-flavour prediction, as observed for $\delta_{CP} = 3.66$.

\subsection{Matter effect in long-baseline experiments}
As active neutrinos propagate from the source to the far detector, they interact with the medium through coherent forward scattering. This is known as the matter effect. These phenomenon modifies the effective Hamiltonian responsible for the flavour evolution of the neutrino states through the introduction of a potential
\begin{equation} 
    H' = H_0 + V = \frac{1}{2E}(\mathcal{M}\mathcal{M}^\dagger + V), 
\end{equation} 
with $\mathcal{M}$ the neutrino mass matrix and $V$ the matter potential matrix. The modification of the effective Hamilton through $V$ alters both the mixing angles and the effective mass-squared differences as neutrinos propagate in matter. All active flavours acquire a neutral-current (NC) potential from $Z$-boson exchange with neutrons while electron neutrinos $\nu_e$ experience an additional charged-current (CC) potential arising from $W$-boson exchange with electrons.  The CC and NC potentials are given by $V_{CC} = \sqrt{2}G_F n_e$ and $V_{NC} = -\frac{1}{\sqrt{2}}G_F n_n$ respectively, where $n_e$ and $n_n$ denote the electron and neutron number densities. For a neutral medium, charge neutrality imposes $n_p = n_e$, and the approximation $n_n \approx n_p$, appropriate for the Earth crust and mantle, yields $V_{CC} = -2V_{NC}$. Since the NC potential is flavour-universal among active neutrinos, it contributes only a global phase to the time evolution and may be removed by a rephasing of the neutrino fields. The effective potential then takes the form 
\begin{equation} 
    V = \textnormal{diag}(2EV_{CC},0,0,-2EV_{NC},-2EV_{NC}), 
\end{equation} 
where $V_{CC} = -2V_{NC} = 3.8 \times 10^{-5} \rho$, with $\rho$ the constant matter density in units of $\textnormal{g}/\textnormal{cm}^3$. For antineutrinos, the potentials reverse sign. The experimental parameters for T2K, NOvA, and DUNE are fixed to baselines $L_\text{T2K} = 295$~km, $L_\text{NOvA} = 810$~km, $L_\text{DUNE} = 1300$~km, with far detector average matter densities $\rho_\text{T2K} \approx 2.6~\text{g}/\text{cm}^3$, $\rho_\text{NOvA} \approx 2.8~\text{g}/\text{cm}^3$, $\rho_\text{DUNE} \approx 2.8~\text{g}/\text{cm}^3$. Adopting constant average densities captures the leading-order matter effect corrections without requiring a position-dependent Earth density profile.  
In the context of quasi-Dirac sterile neutrinos, matter effects introduce an important modification to the oscillation phenomenology \cite{Anamiati:2017rxw, Anamiati:2019maf}. For splittings $\Delta m_{54}^2 \ll \Delta m_{31}^2$, the vacuum oscillation length $L_{54} = 4\pi E_\nu/\Delta m_{54}^2$ exceeds the experimental baseline and the associated phase remains negligibly small in vacuum \cite{Anamiati:2017rxw}. The matter potential shifts the eigenvalues of the effective $5\times5$ Hamiltonian by corrections of order $V_{CC}/\Delta m_{54}^2$, inducing a non-trivial phase between the quasi-Dirac partners $\nu_4$ and $\nu_5$, with sign determined by $\text{sgn}(\Delta m_{54}^2)$ and magnitude scaling linearly with $n_e$ \cite{Anamiati:2017rxw, Donini:2011jh}. As shown in Fig.~\ref{NO_IO}, the splitting $\Delta m_{54}^2$ is taken comparable in magnitude to $\Delta m_{21}^2$, precisely the regime in which such matter-induced modifications to the active-sterile level structure become non-negligible at long baselines \cite{Anamiati:2019maf}. NOvA and in particular DUNE are thus well-suited to the quasi-Dirac mass splitting $\Delta m_{54}^2$ through distortions in the energy-dependent $\nu_\mu \to \nu_e$ appearance and $\nu_\mu$ disappearances spectra, with projected sensitivity extending to splittings $\Delta m_{54}^2 \sim 10^{-5}$--$10^{-4}~\text{eV}^2$ \cite{Anamiati:2019maf}.

\subsection{Sterile neutrino decay lengths}

Sterile neutrinos are unstable and decay to SM particles via charged- and neutral-current processes, suppressed via the active-sterile mixing $\abs{U_{\alpha N}}^2\approx\theta_{sa}^2$, valid for $\theta_{sa}\ll1$ \cite{Bolton_2025}.  The sterile neutrino decays proceed through off-shell $W^\pm$ and $Z$ bosons. For the small mixing angles of interest, the resulting decay lengths are of the order of millions of kilometers and larger, making these states well suited for searches at fixed-target experiments. Given the masses quoted in Table \ref{NO_IO}, hadronic and mesonic final states are kinematically inaccessible, and decays to $\mu$ and $\tau$ are forbidden. The decay channels therefore reduce to $N \to \nu_\alpha e^+ e^-$ and $N \to \nu_\alpha \bar{\nu}_\beta \nu_\beta$. For the decay mode $N \to \nu_\alpha e^+ e^-$, both CC and NC diagrams contribute when $\alpha = e$, leading to an interference between the two, while for $\alpha \neq e$ only the NC diagram contributes. For the purely leptonic channel, since $\Gamma(N \to \nu_\alpha \bar{\nu}_\beta \nu_\beta)$ is independent of $\beta$, the sum over $\beta = e, \mu, \tau$ contributes an overall factor of 3. The total decay width is therefore
\begin{equation}
    \Gamma_N = \sum_{\alpha=e}^{\tau} \Gamma(N \to \nu_\alpha e^+ e^-) 
    + 3\sum_{\alpha=e}^{\tau} \Gamma(N \to \nu_\alpha \bar{\nu}_e \nu_e).
\end{equation}

\begin{figure}[t!]
    \centering
    \includegraphics[width=0.6\linewidth]{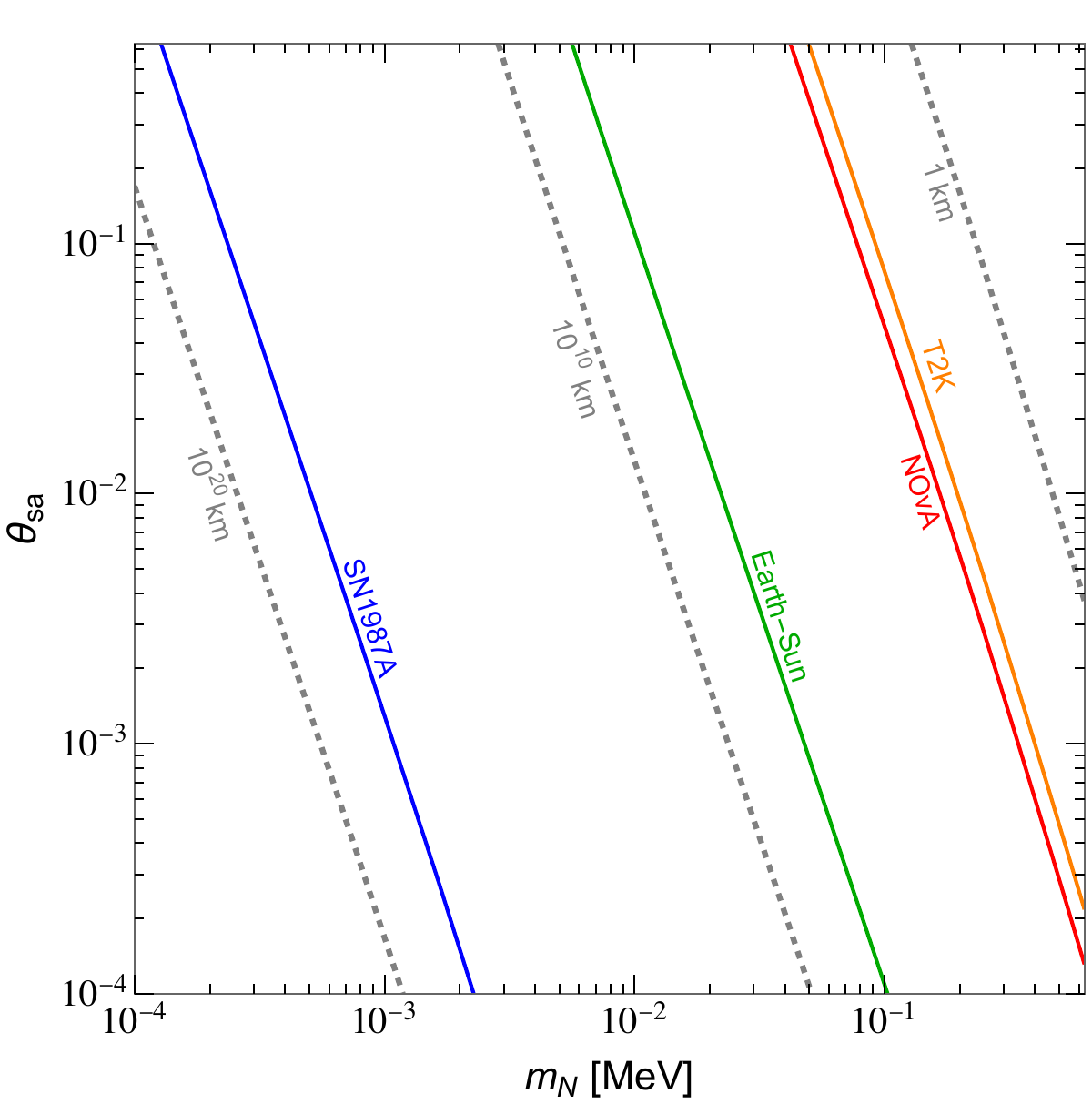}
    \caption{Lab-frame decay length $L_{\rm lab}$ of a sterile neutrino with energy $E_N = 1$ GeV as a function of mass $m_N$ and mixing angle $\theta_{sa}$. The solid lines indicate the baselines of T2K ($295$ km), NO$\nu$A ($810$ km), the Earth-Sun distance ($1.496\times 10^{8}$ km), and SN1987A ($1.68\times 10^{18}$ km), as indicated.}\label{figure::decays}
\end{figure}

Below we provide expressions for each channel; a more detailed analysis can be found in \cite{Bondarenko_2018}. The decay width for $N \to \nu_\alpha e^+ e^-$, including CC and NC contributions and their interference, is
\begin{align}
    \Gamma(N \to \nu_\alpha e^+ e^-) &= \frac{G_F^2 m_N^5}{192\pi^3}\theta_{sa}^2
    \bigg[C_1\Big((1-14x^2-2x^4-12x^6)\sqrt{1-4x^2} + 12x^4(x^4-1)L(x)\Big) \nonumber\\
    &\quad + 4C_2\Big(x^2(2+10x^2-12x^4)\sqrt{1-4x^2} + 6x^4(1-2x^2+2x^4)L(x)\Big)\bigg],
\end{align}
where $x = m_e/m_N$ and
\begin{equation}
    L(x) = \log\!\left[\frac{1-3x^2-(1-x^2)\sqrt{1-4x^2}}{x^2(1+\sqrt{1-4x^2})}\right].
\end{equation}
The coefficients $C_1$ and $C_2$ are given by
\begin{align}
    C_1 = \tfrac{1}{4}(1\pm4\sin^2\theta_W + 8\sin^4\theta_W), \qquad
    C_2 = \tfrac{1}{2}\sin^2\theta_W(2\sin^2\theta_W\pm1)
\end{align}
where the upper sign applies for $\alpha = e$ and the lower sign for $\alpha \neq e$. For the purely leptonic channel, only NC diagrams contribute and the relevant decay width is
\begin{equation}
    \Gamma(N \to \nu_\alpha \bar{\nu}_e \nu_e) = 
    (1+\delta_{\alpha e})\frac{G_F^2 m_N^5}{768\pi^3}\theta_{sa}^2.
\end{equation}
The decay length in the lab frame is $L_\text{lab} = \beta\gamma/\Gamma_N$, and depends strongly on both $\theta_{sa}$ and $m_N$, as illustrated in Fig.~\ref{figure::decays}. The scaling $\Gamma_N \propto \theta_{sa}^2 m_N^5$ implies that small mixing angles and light masses yield characteristically long-lived states. For the parameter values in Table~\ref{table_params}, the resulting decay lengths substantially exceed any terrestrial detector baseline, and the analysis therefore operates within the long-lived regime in which neutrino oscillations are observable. The same applies to solar and supernova neutrinos, for which the relevant propagation distances are of order $10^{11}$ m and kiloparsecs respectively, rendering prompt decay negligible across the entirety of the parameter space considered.

\section{Simulating quasi-Dirac neutrino oscillations} \label{sec::simulation}
Making the observation that $\Delta L = 4\pi E/\Delta m_{ij}^2$ represents the baseline for a full oscillation period, we note that the introduction of a small mass splitting is equivalent to introducing new, large oscillation periods $\Delta L$.  This implies that experiments with large baselines are best suited to constrain $\Delta m_{54}^2 = 10^{-3}~\text{eV}^2$. In addition, they enhance sensitivity to the amplitude of new oscillation modes governed by the mixing angles $\theta_{ij}$. T2K, NOvA, and DUNE operate with beams with a flux of predominantly $\nu_\mu$ $(\overline{\nu_\mu})$ when operating in (anti)neutrino mode. These experiments are designed to study the disappearance of $\nu_\mu$ and and the appearance of $\nu_e$. Using GLoBES (General Long Baseline Experiment Simulator) \cite{Huber:2004ka}, we determine the expected event rates in the disappearance and appearance channels of NOvA, T2K and DUNE given a set of $3\nu$ and $5\nu$ neutrino oscillation parameters. We use the existing best-fit values of the $3\nu$ PMNS matrix parameters and mass splittings from NuFit \cite{Esteban:2024eli}, and $\theta_{sa}$ and $\Delta m_{54}^2$ within the range outlined by Table \ref{table_params}. The large mass splitting $\Delta m_{41}^2$ is kept fixed, as its associated oscillation length is sufficiently short that the corresponding oscillations are averaged out at the baselines considered. We compare the expected event rates in the $5\nu$ paradigm against the expected $3\nu$ event rates and construct a $\chi^2$ test statistic to assess the sensitivity of the experiment to the $5\nu$ model.

\subsection{Simulating NOvA, T2K and DUNE using GLoBES}
\label{sec:modelling}
In this work, we use GLoBES (General Long Baseline Experiment Simulator) \cite{Huber:2007ji} to model neutrino oscillations and event rates in long-baseline experiments. GLoBES enables the simulation of detector setups, fluxes, cross sections, and systematics via its Abstract Experiment Definition Language (AEDL) \cite{Huber:2004ka}. It computes oscillation probabilities and convolves them with experimental inputs to generate event spectra and to evaluate $\chi^2$ statistics for hypothesis comparisons. The framework also allows the customisation of oscillation probabilities and statistics, making it suitable for the exploration of BSM scenarios.
Specifically, we use GLoBES to assess the sensitivity of the NO$\nu$A, T2K and DUNE experiments to quasi-Dirac sterile neutrinos by comparing our five-flavour oscillation model with the standard three-flavour case, examining deviations in observables and mapping the relevant parameter space for sterile-active mixing.

\begin{table}[t!]
\centering
\begin{tabular}{c|c|c|c}
\hline
Experimental Configuration & NO$\nu$A~\cite{NOvA:2025tmb} & T2K~\cite{Fechner:2006koa} & DUNE~\cite{DUNE:2021cuw} \\ \hline
Baseline [km] & 810 & 295 & 1285 \\
Fiducial target mass [kt] & 14 & 22.5 & 40 \\
Energy Range [GeV] & 0-5 & 0-2 & 0-8 \\
 POTs $\nu$ & $26.6\times 10^{20}$ & $20 \times 10^{20}$ & $71.5\times 10^{20}$\\
POTs $\bar\nu$ & $12.5\times 10^{20}$ & $18\times10^{20}$  & $71.5\times 10^{20}$ \\
Run Time [yrs] $\nu$ & 6.6  & 5 & 6.5   \\
Run Time [yrs] $\bar\nu$ &3 & 5 & 6.5 
\\\hline
\end{tabular}
\caption{Detector configurations of the experiments used in the simulation of event rates and the sensitivity analysis.}
\label{tab:exp_configs}
\end{table}
\textbf{NOvA} -- NO$\nu$A is a long-baseline neutrino oscillation experiment operating with a baseline of $810$ km between Fermilab and Ash River, Minnesota, with a peak beam energy of $\sim 2$ GeV. It studies $\nu_\mu$ disappearance and $\nu_e$ appearance using a primarily $\nu_\mu$ ($\bar{\nu}_\mu$) beam in neutrino (antineutrino) mode, and has accumulated substantial exposure in both channels. Our simulation of NOvA is designed to study the expected event rates in the $\nu_\mu$ disappearance channel and the $\nu_e$ appearance channel with neutrino energies ranging from 0 to 5 GeV over 19 bins. We incorporate realistic inputs for the NOvA beam flux, energy resolution, cross section and detector efficiencies. Specifically, the neutrino interaction cross section data is taken from \cite{NOvA:2021eqi,NOvA:2024rov} for a total exposure of $26.61 \times 10^{20}$ protons-on-target (POTs) in Forward Horn Current (FHC) mode and $12.50 \times 10^{20}$ POTs in Reverse Horn Current (RHC) mode, accumulated between 2014 and 2024 in the NuMI beam.

\begin{figure}[t!]
    \centering
    \includegraphics[width=0.49\linewidth]{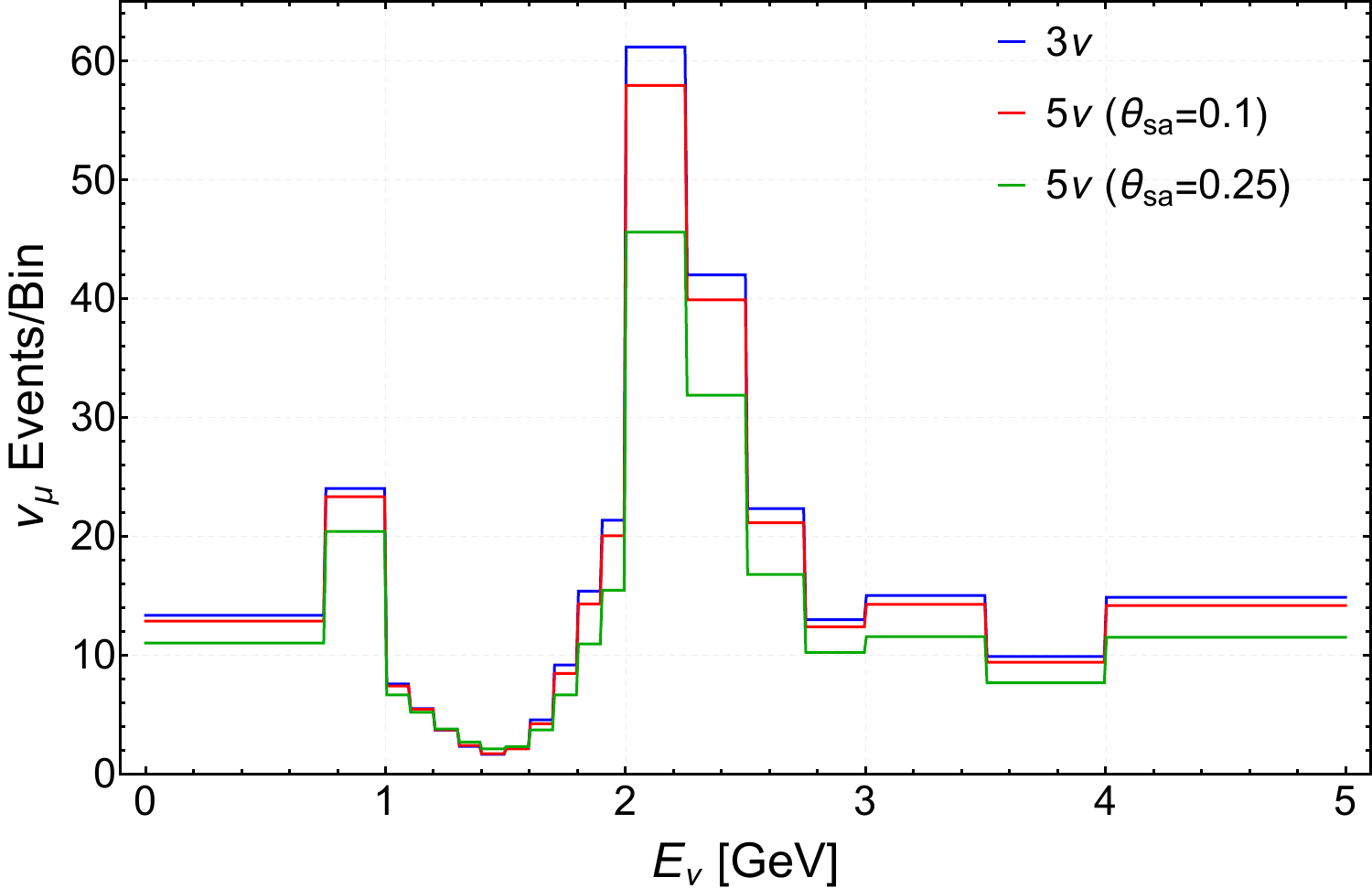}
    \includegraphics[width=0.49\linewidth]{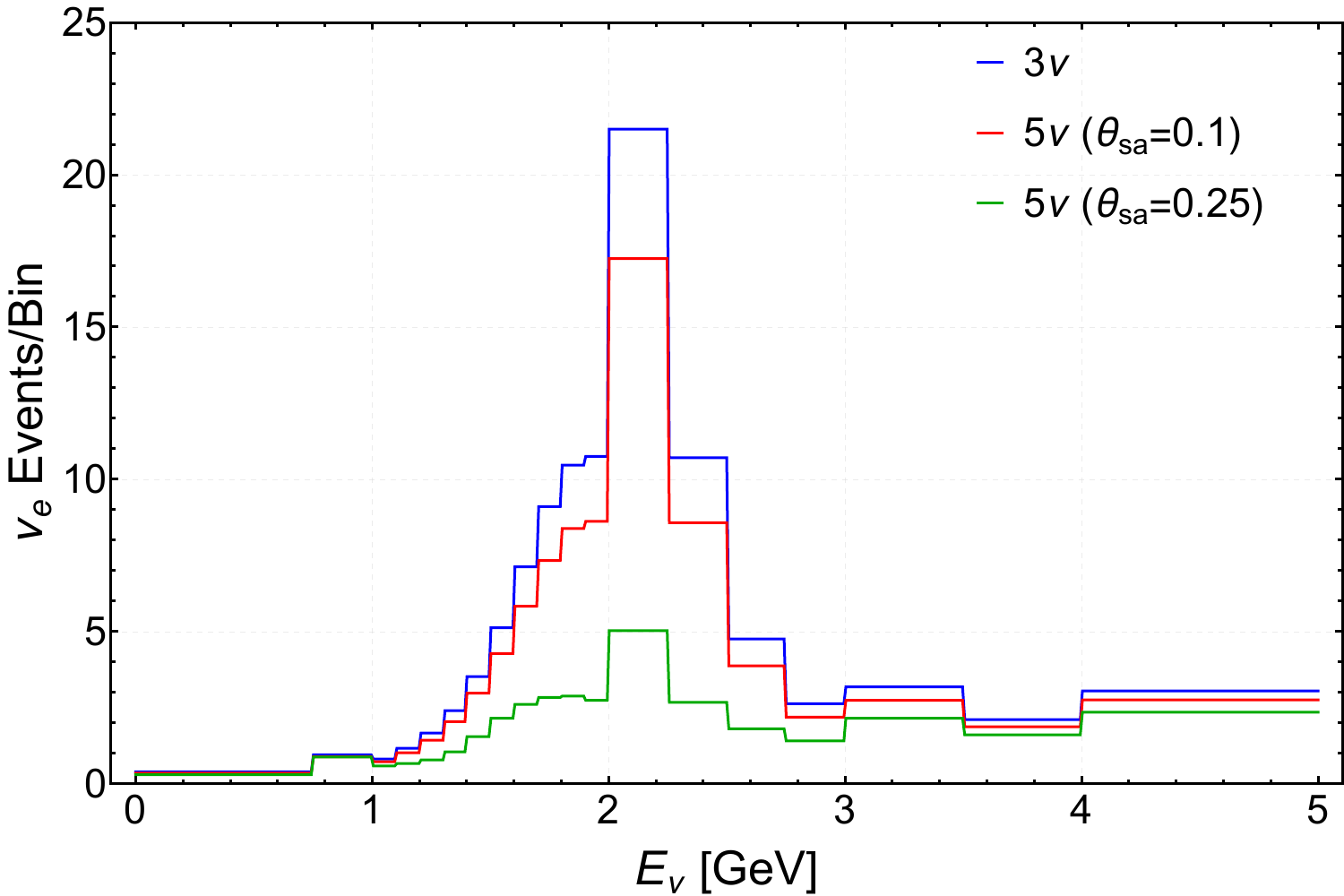}
    \caption{Expected number of $\nu_\mu$ disappearance (left) and $\nu_e$ appearance (right) events at the NOvA experiment, comparing $3\nu$ (blue) to $5\nu$ for a mass squared splitting of $\Delta m^2_{54}=0.01~\textnormal{eV}^2$ with different mixing angles $\theta_{sa}=0.1$ (red) and $\theta_{sa}=0.25$ (green)}
\label{fig:nova-comparison}
\end{figure}

Fig.~\ref{fig:nova-comparison} illustrates the expected event rates in NOvA in the $\nu_\mu \to \nu_\mu$ and in the $\nu_\mu \to \nu_e$ channels given specific sets of $\theta_{sa}$ and $\Delta m_{54}^2$. We notice that an enhanced sensitivity in the appearance channel relative to disappearance arises because the $\nu_\mu \to \nu_e$ transition amplitude receives direct contributions from the active-sterile mixing angles $\theta_{14}$, $\theta_{15}$, $\theta_{24}$, $\theta_{25}$ at leading order, opening new oscillation pathways that are absent in the $3\nu$ case. The $\nu_\mu \to \nu_\mu$ survival probability, by contrast, is primarily governed by $\theta_{23}$ and $\Delta m_{31}^2$, with the sterile mixing entering only as a subleading correction through the unitarity deficit $1 - \sum_{i=1}^{5}|U_{\mu i}|^2 = 0$. Consequently, the active-sterile mixing $\theta_{sa}$ produces a more pronounced distortion in the appearance spectrum than in the disappearance spectrum at the same parameter values.

\begin{figure}[t!]
    \centering
    \includegraphics[width=0.48\linewidth]{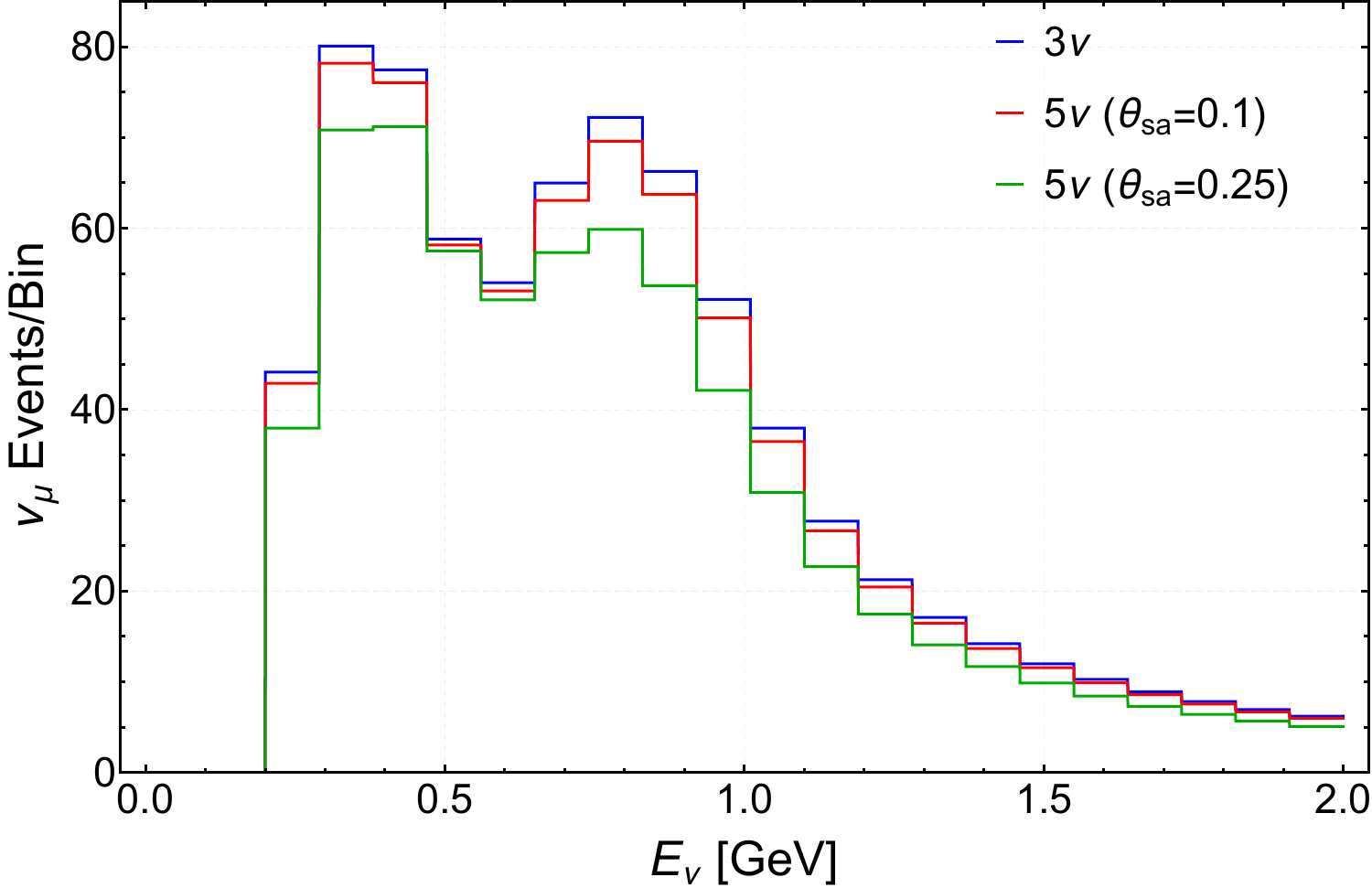}
    \includegraphics[width=0.48\linewidth]{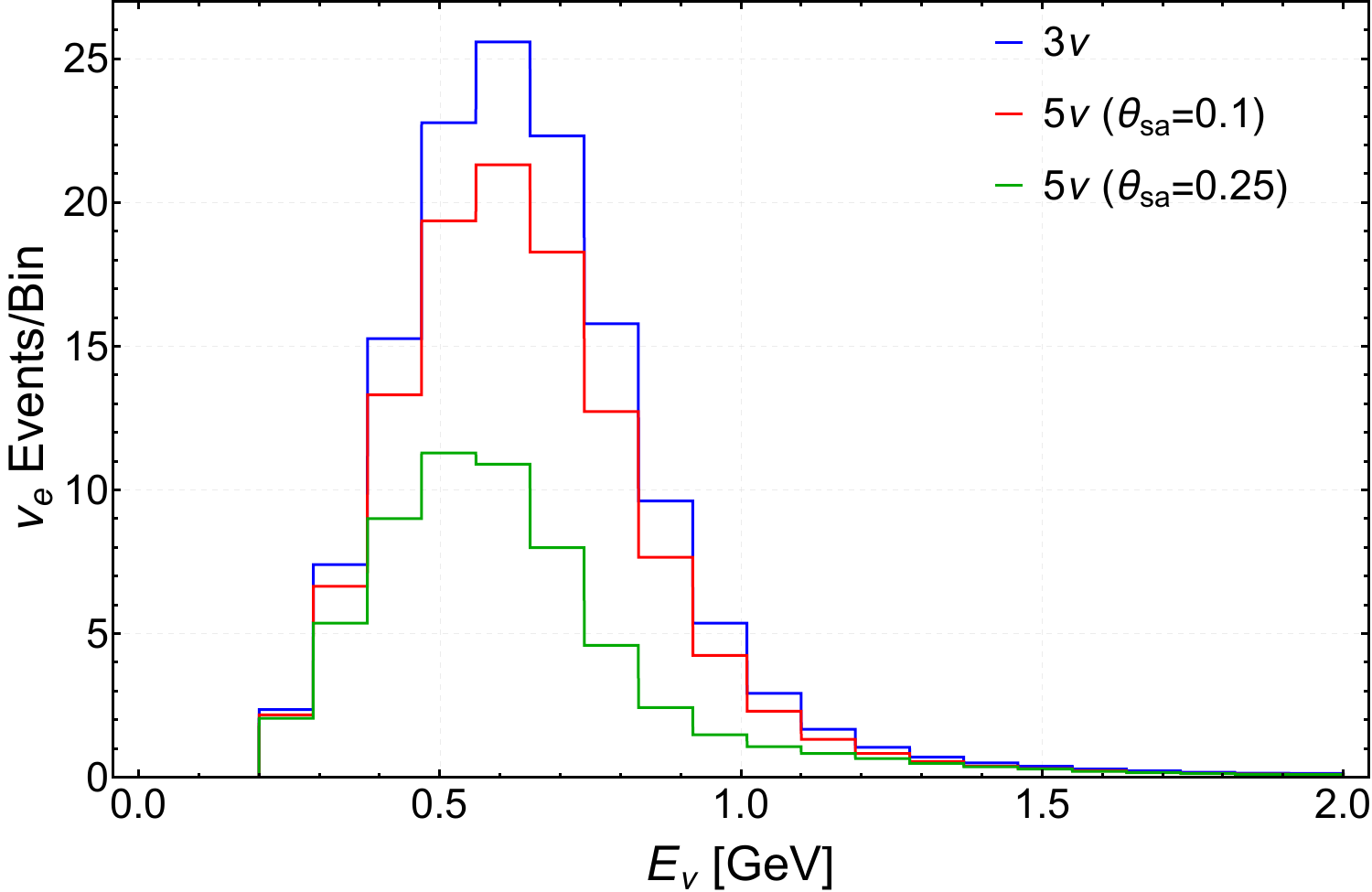}
    \caption{As Fig.~\ref{fig:nova-comparison} but at the T2K experiment.}
\label{fig:t2k-comparison}
\end{figure}

\textbf{T2K} -- T2K is a long-baseline neutrino oscillation experiment operating with a baseline of $295$ km between J-PARC and the Super-Kamiokande detector in Japan, with a peak beam energy of $\sim 0.6$ GeV. It similarly studies $\nu_\mu$ disappearance and $\nu_e$ appearance, and provides complementary sensitivity to NO$\nu$A owing to its different $L/E$ profile and detector technology. We simulate T2K in a similar fashion to NOvA, by simulating the expected event rates in the $\nu_\mu \to \nu_e$ and in the $\nu_\mu \to \nu_\mu$ channels for $3\nu$ and $5\nu$ oscillations. In this case, the event rates are plotted over 30 bins ranging from 0 to 2 GeV. Specifically, the neutrino cross section is taken from \cite{Campagne_2007} for a total exposure of $20 \times 10^{20}$ POTs over a 5 year period from January 2010. Fig.~\ref{fig:t2k-comparison} illustrates the expected event rates at T2K in the $\nu_\mu$ disappearance and in the $\nu_e$ appearance channel for given values of $\theta_{sa}$ and $\Delta m_{54}^2$. As with NOvA, the discrepancy between the $3\nu$ and $5\nu$ predictions is more pronounced in the appearance channel than in the disappearance channel, for the reasons discussed above. However, T2K's narrower energy range of $0$--$2$ GeV compared to NOvA's $0$--$5$ GeV means the appearance spectrum is probed over a more restricted window, which limits sensitivity to values of $\Delta m_{54}^2$ whose associated oscillation length falls outside this range. The different $L/E$ profile further means that the distortion in the appearance spectrum manifests at different neutrino energies to NOvA, providing complementary rather than redundant constraints on the quasi-Dirac parameter space.

\textbf{DUNE} -- DUNE is a future neutrino experiment that is expected to start taking data in 2029. With a near detector at the Sanford Underground Research Facility and a far detector at Fermilab, neutrinos will travel a distance of 1300 km. Our simulation of DUNE includes the expected event rates in the $\nu_\mu$ disappearance and $\nu_e$ appearance channels with neutrino energies ranging from 0 to 8 GeV over 64 bins. As this is a future experiment, we base our analysis on the predicted experimental configurations outlined in \cite{DUNE:2021cuw} for a total runtime of 6.5 years in both neutrino and antineutrino mode. 

\begin{figure}[t!]
    \centering
    \includegraphics[width=0.48\linewidth]{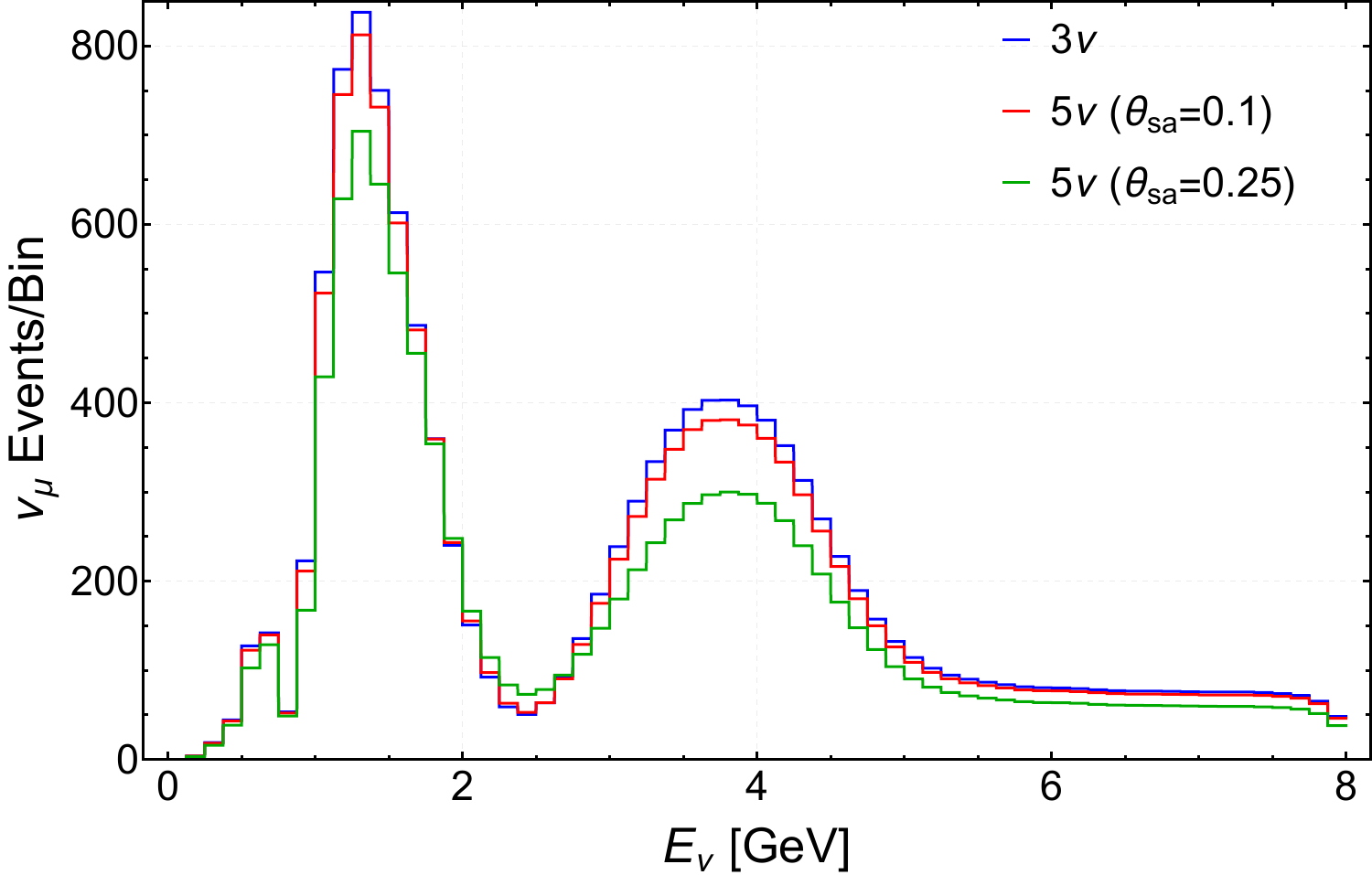}
    \includegraphics[width=0.48\linewidth]{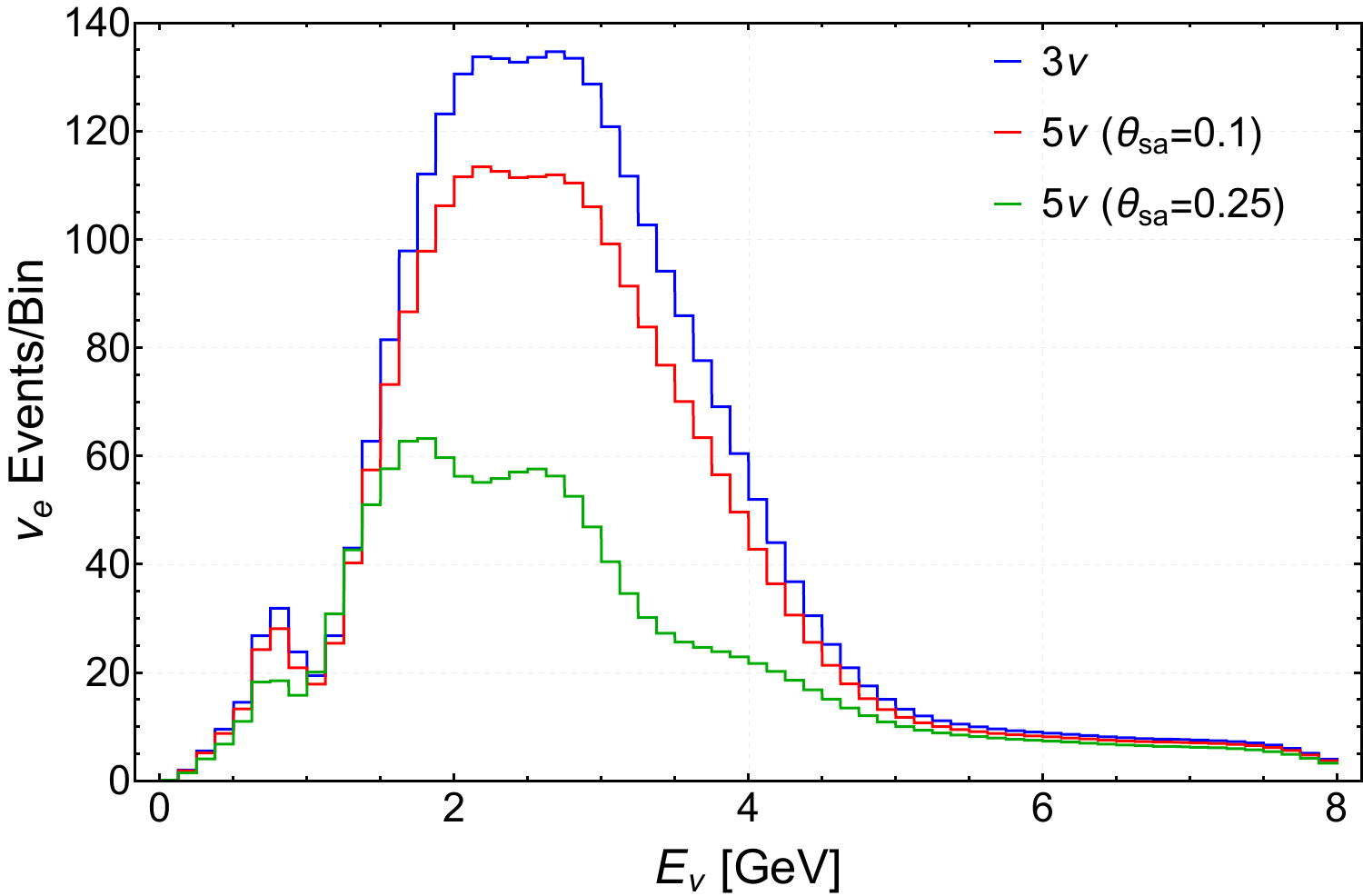}
    \caption{As Fig.~\ref{fig:nova-comparison} but at the projected DUNE experiment.}
\label{fig:dune-comparison}
\end{figure}

Fig.~\ref{fig:dune-comparison} illustrates the expected event rates at DUNE in the $\nu_\mu \to \nu_\mu$ and  $\nu_\mu \to \nu_e$ channels. We can see that this experiment is far more sensitive to $\Delta m_{54}^2$ and particularly to $\theta_{sa}$ compared to NO$\nu$A and T2K.  This enhanced sensitivity is primarily a consequence of DUNE's significantly larger detector volume and higher beam intensity, which together yield a substantially greater POT exposure and hence event statistics, particularly in the appearance channel where the sterile mixing effects are most pronounced.

\subsection{Sensitivity to model parameters}

We perform a $\chi^2$ analysis to evaluate the sensitivity of the NOvA experiment to sterile neutrino signatures by comparing expected event rates under the standard $3\nu$ and extended $5\nu$ oscillation models. For $k$ energy bins, the test statistic is defined as
\begin{equation}\label{chi_squared}
    \chi^2=\sum_k\left(\mu_k-n_k+n_k\ln{\left(\frac{n_k}{\mu_k}\right)}\right) +  \sum_k \left( \frac{x_k - x_{k,\text{center}}}{\sigma_k} \right)^2,
\end{equation}
where $n_k$ and $\mu_k$ denote the predicted event rates in bin $k$ for the $3\nu$ and $5\nu$ cases, respectively. The second term represents a Gaussian prior on a set of nuisance parameters $x_k$, which typically models uncertainties like signal normalization as described below. This prior is added to $\chi^2$ and serves to penalize large deviations from the central values $x_{k\text{,center}}$. To properly account for the corresponding systematic uncertainties $\sigma_k$, we minimize the $\chi^2$ over the nuisance parameters $x_k$. This analysis is implemented in GLoBES, where we choose the $3\nu$ model as the null hypothesis to probe the $5\nu$ model. The resulting $\chi^2$ values quantify the sensitivity of the experiment to differences between these models. 

We conduct this analysis through a direct comparison of the predicted event rates in the $3\nu$ and $5\nu$ oscillation frameworks. Realistic uncertainties associated with flux normalization, detector response, cross sections, and background estimation are incorporated as nuisance parameters, which are varied during the $\chi^2$ minimization procedure as shown in Eq.~(\ref{chi_squared}). This allows part of the discrepancy between the $3\nu$ and $5\nu$ predictions to be absorbed by systematic shifts, reducing the resulting $\chi^2$ values and yielding more conservative sensitivity estimates than a purely statistical treatment. The $\chi^2$ is computed as a function of $\Delta m_{54}^2$ and $\theta_{sa}$ using a spectrum-wide analysis, comparing event rate spectra across all energy bins. The degree of sensitivity degradation varies across experiments according to their respective systematic uncertainty budgets and the regions of parameter space where the $3\nu$ and $5\nu$ predictions are most degenerate, but is typically around 0.02 radians in the active-sterile mixing angle. Nevertheless, this treatment is essential for a rigorous and experimentally faithful analysis, and all sensitivity results presented in this work include full systematic uncertainties.
The analysis incorporates systematic uncertainties tailored to each experimental setup. For NOvA, these include a 5\% uncertainty on signal normalization, 2.5\% on signal calibration, 10\% on background normalization, and 2.5\% on background calibration. The T2K uncertainties are set at 2\% on signal normalization, 1\% on signal calibration, and 5\% on both background normalization and calibration for the $\nu_e$ and $\bar{\nu}e$ channels; for the $\nu\mu$ and $\bar{\nu}_\mu$ channels, all uncertainties are reduced to 0.1\%. The treatment of systematic uncertainties in DUNE is more involved, owing to its joint near and far detector analysis. Signal uncertainties are set at 2\% for $\nu_e$ and 5\% for $\nu_\mu$, with the same values applied to their respective backgrounds. Both signal and background normalizations enter the fit as nuisance parameters, constrained through penalty terms within the GLoBES framework, and are marginalized over alongside energy calibration uncertainties. This approach provides a more comprehensive treatment of the detector's systematic effects than a far-detector-only analysis would afford.

To prevent numerical aliasing, which can arise from rapidly oscillating terms in the calculated neutrino oscillation probability, we utilize a low-pass Gaussian filter provided within the GLoBES framework. Without this filter, these high-frequency components can introduce non-physical artifacts into the simulation. The filtering is achieved by incorporating a Gaussian damping term into the probability calculation, which modifies the standard formula in Eq. \eqref{oscillation_eqn} as
\begin{equation}
    \label{eq:gaussian_prob}
    P_{\alpha \beta}(E) = \sum_{ij} U_{\alpha j}U^*_{\beta j}U^*_{\alpha i}U_{\beta i} e^{-i\phi_{ij}} \times e^{-\frac{\sigma_e^2 \phi_{ij}^2}{E^2}},
\end{equation}
which includes a Gaussian low frequency filter, where $\sigma_e$ defines the width of the energy smearing. The high-frequency oscillations originating from the presence of sterile neutrinos with oscillation lengths much shorter than the baseline get suppressed by the filtering factor making the event rate calculations based on the averaged oscillations rather than the high-frequency oscillations. This makes the sensitivity more uniform in terms of calculations numerically and as a consequence the curvature of the sensitivity becomes smoother.

\section{Results} 
\label{sec::results}
The sensitivity to sterile neutrino effects is quantified by the region in the $(\theta_{sa}, \Delta m^2_{54})$ plane where the $5\nu$ and $3\nu$ predictions become distinguishable. Since NO$\nu$A, T2K and DUNE each operate in both neutrino-mode (FHC) and antineutrino-mode (RHC), we present the $\chi^2$ sensitivity contours for all channels and compare across the three experiments.The sensitivity to sterile neutrino effects is quantified by the region in the $(\theta_{sa}, \Delta m^2_{54})$ plane where the $5\nu$ and $3\nu$ predictions become distinguishable. Since NO$\nu$A, T2K and DUNE each operate in both neutrino-mode (FHC) and antineutrino-mode (RHC), we present the $\chi^2$ sensitivity contours for all channels and compare across the three experiments. The remaining standard oscillation parameters are fixed to their NuFit 6.0 best-fit values \cite{Esteban:2024eli}, with $\Delta m^2_{41}$ fixed at $10~\text{eV}^2$; the full parameter set is given in Table~\ref{table_params}.

\begin{figure}[t!]
    \centering
    \includegraphics[width=1\linewidth]{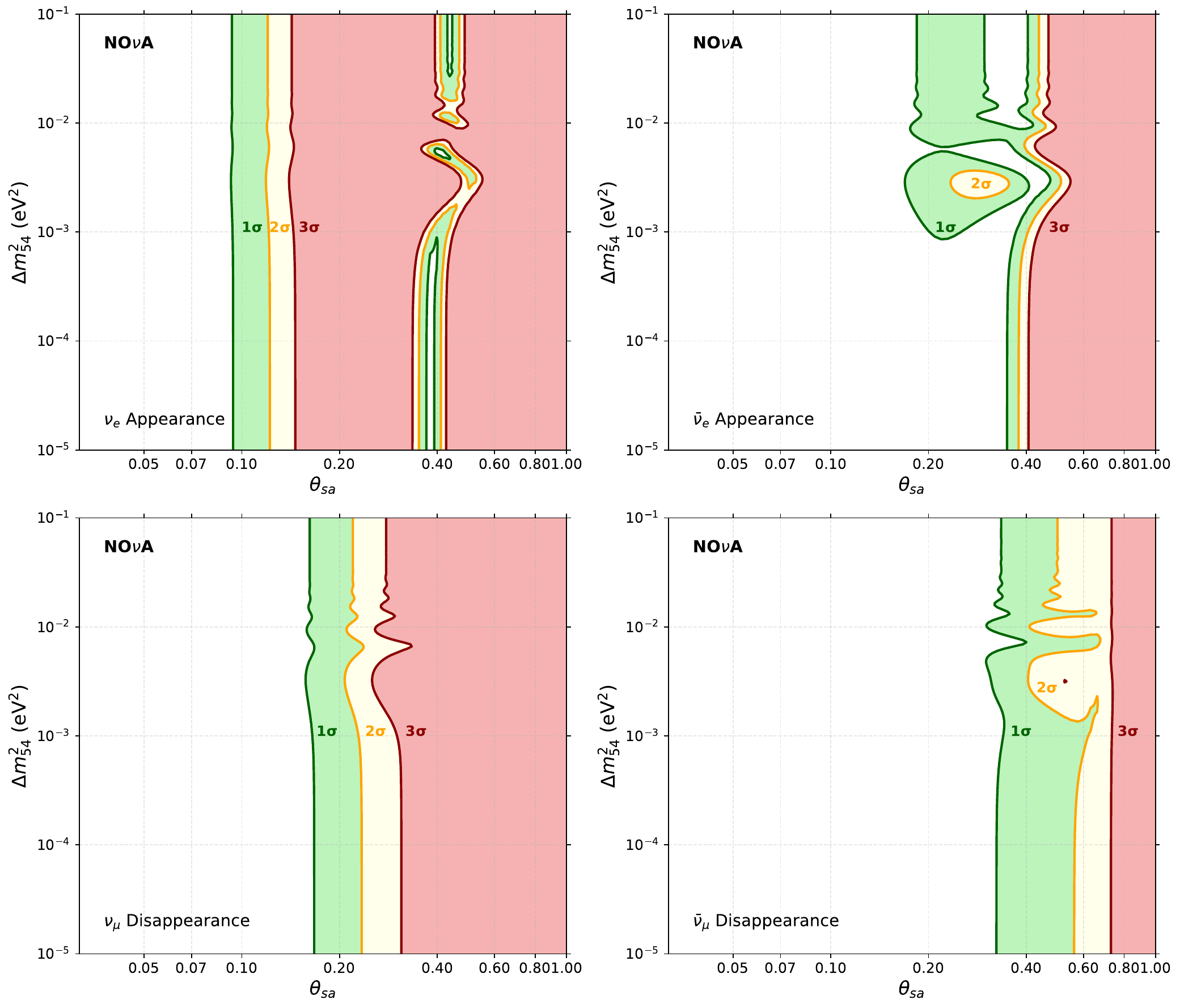}
    \caption{$\chi^2$ sensitivity contours of NO$\nu$A in the plane of mass square splitting $\Delta m^2_{54}$ and sterile mixing $\theta_{sa}$ for the event channels of $\nu_e$ appearance (top left), $\bar{\nu_e}$ appearance (top right), $\nu_\mu$ disappearance (bottom left) and $\bar{\nu_\mu}$ disappearance (bottom right). The contour regions represent the confidence levels as the white area is the allowed region within 1$\sigma$ (68\% C.L.), green corresponds to the 1$\sigma$–2$\sigma$ range (68\%–95\% C.L.), yellow to the 2$\sigma$–3$\sigma$ range (95\%–99.7\% C.L.), and red indicates regions excluded at greater than 3$\sigma$. The $\chi^2$ sensitivity is calculated with systematics included and filtering.}
    \label{fig:nova_4_channels}
\end{figure}
\begin{figure}[t!]
    \centering
    \includegraphics[width=1\linewidth]{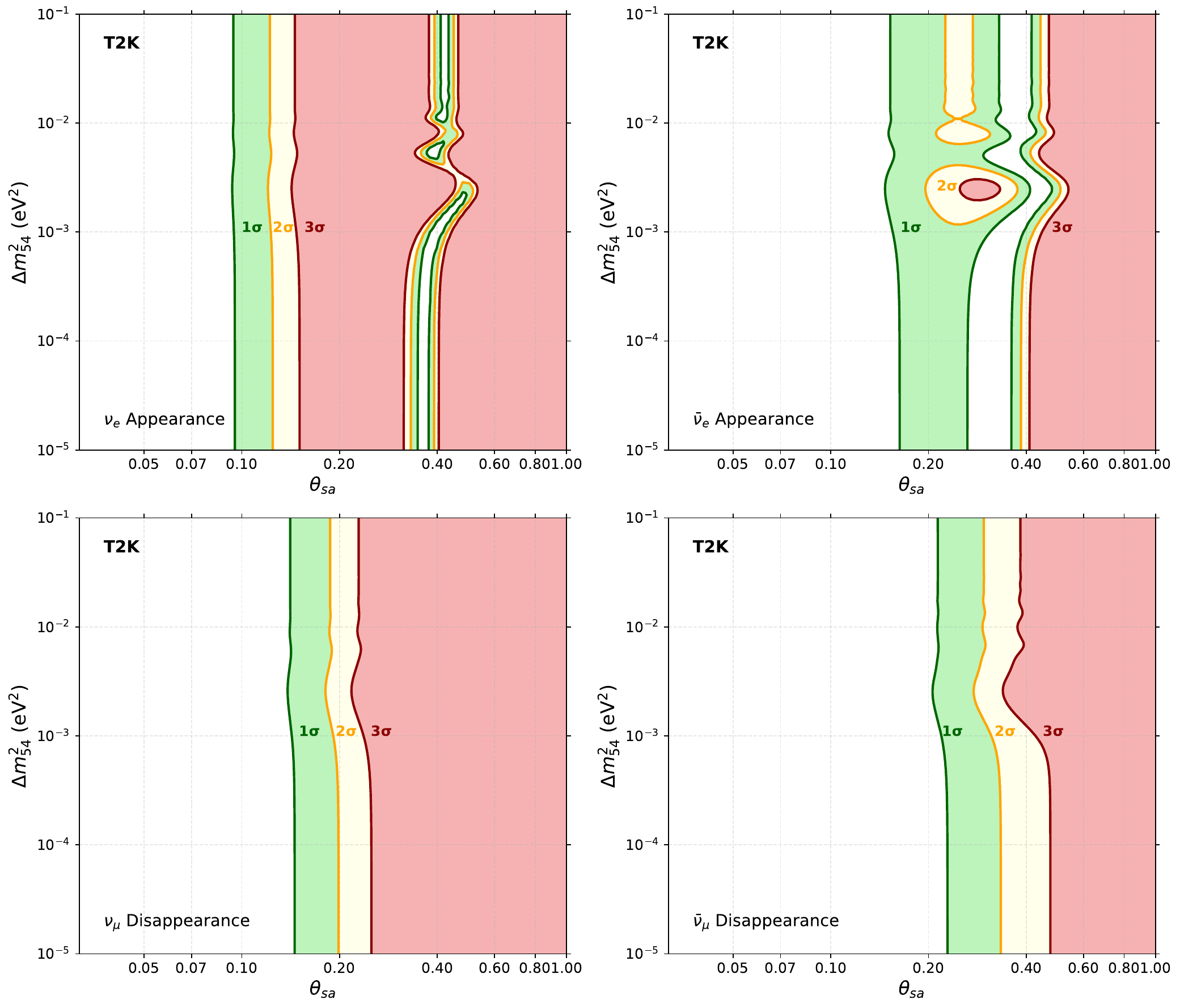}
    \caption{$\chi^2$ sensitivity contours for T2K in the $\Delta m^2_{54}$–$\theta_{sa}$ plane for the same four event channels and confidence level conventions as Fig.~\ref{fig:nova_4_channels}.}
    \label{fig:t2k_4_channels}
\end{figure}
\begin{figure}[t!]
    \centering
    \includegraphics[width=1\linewidth]{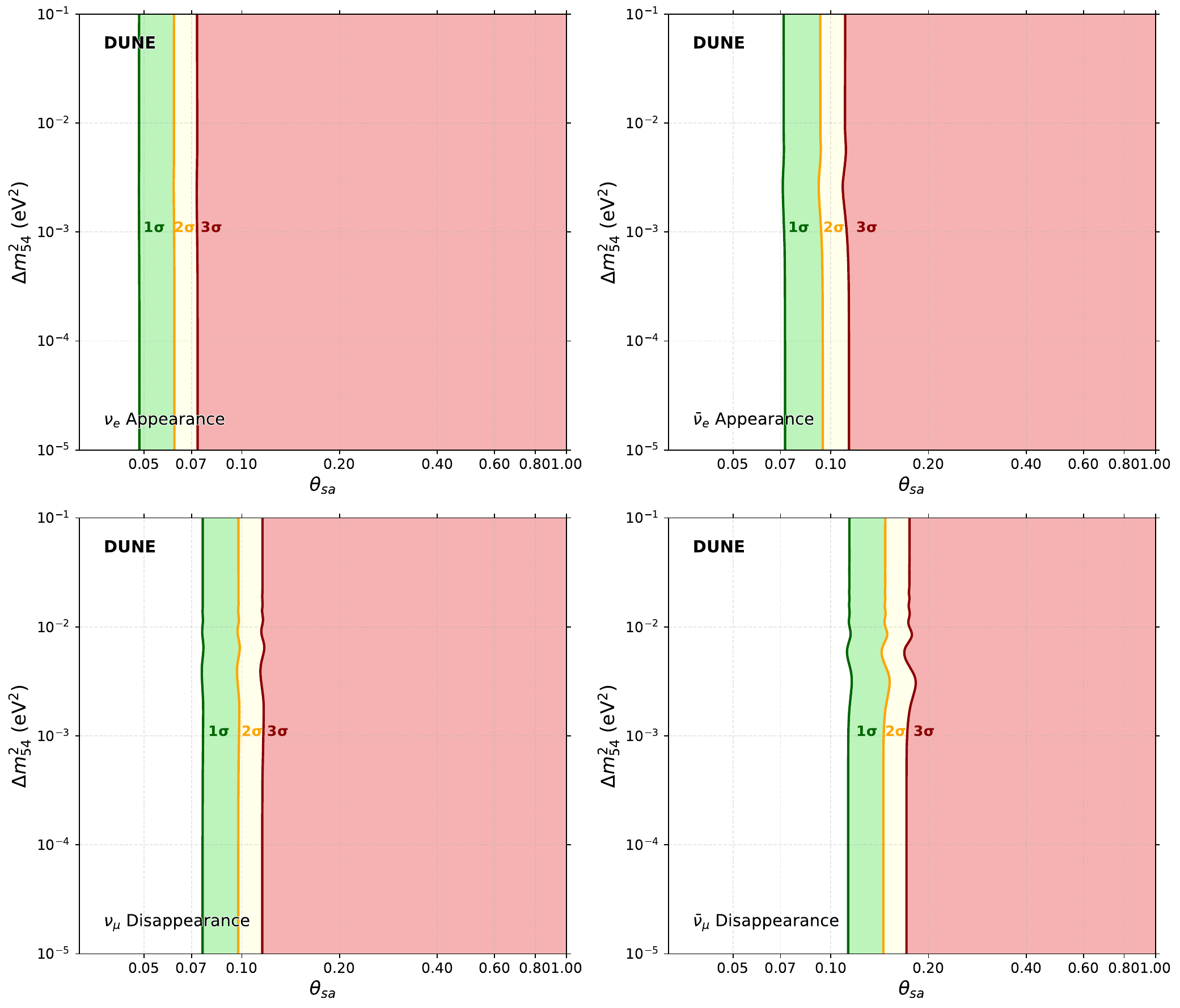}
    \caption{$\chi^2$ sensitivity contours for DUNE in the $\Delta m^2_{54}$–$\theta_{sa}$ plane for the same four event channels and confidence level conventions as Fig.~\ref{fig:nova_4_channels}.}
    \label{fig:DUNE_4_channels}
\end{figure}
\textbf{NO$\nu$A} -- The $\chi^2$ sensitivity contours for NO$\nu$A are shown in Fig.~\ref{fig:nova_4_channels}. Across all channels, a common trend is observed: the experimental sensitivity is greater at larger values of $\Delta m^2_{54}$, while in the low mass-splitting regime, the two sterile states become nearly degenerate and effectively behave as a single Dirac neutrino, rendering their individual contributions indistinguishable. In the higher mass-splitting regime, oscillatory structures appear across the channels, arising when $\Delta m^2_{54}$ is of an order such that the sterile-induced oscillation length becomes comparable to the experimental baseline. The neutrino channels exhibit higher sensitivity than their anti-neutrino counterparts, owing to the longer operation of NO$\nu$A in FHC mode and the correspondingly larger accumulated exposure. Overall, NO$\nu$A is sensitive to sterile-active mixing down to $\theta_{sa} \approx 0.1$. A notable feature appears in the $\bar{\nu}_e$ appearance and $\bar{\nu}_\mu$ disappearance channels, where the sensitivity is visibly enhanced in the vicinity of $\Delta m^2_{54} \sim 2.4 \times 10^{-3}~\text{eV}^2$. This value coincides with the atmospheric mass splitting $\Delta m^2_{31}$, and the enhancement arises from a resonance effect: when the sterile-induced oscillation length becomes comparable to that of the standard atmospheric oscillation, the two interfere constructively, producing enhanced spectral distortions and a corresponding increase in the $\chi^2$ value. This interference, however, also introduces parameter degeneracies between $\Delta m^2_{54}$ and the atmospheric oscillation parameters ($\Delta m^2_{31}$, $\theta_{23}$), complicating their individual interpretation. A second structural feature is the emergence of low-$\chi^2$ valleys near $\theta_{sa} \approx 0.4$, which is of the same order as the standard atmospheric mixing angle $\theta_{23}$. At this value, the sterile contribution to the oscillation amplitude becomes comparable to the active-active contribution, giving rise to a partial cancellation in the $\chi^2$ landscape. For specific combinations of $\theta_{sa}$ and $\Delta m^2_{54}$, the $5\nu$ model can therefore reproduce the $3\nu$ spectral shape closely enough to evade detection, manifesting as a degeneracy in the sensitivity contours.

\textbf{T2K} -- The $\chi^2$ sensitivity contours for T2K are shown in Fig.~\ref{fig:t2k_4_channels}. The qualitative features observed in the NO$\nu$A contours, the atmospheric resonance enhancement near $\Delta m^2_{54} \sim \Delta m^2_{31}$, the low-$\chi^2$ degeneracy valleys near $\theta_{sa} \approx 0.4$ and the suppressed sensitivity at low mass splittings, are reproduced in T2K at the same parameter values. The recurrence of these features across two independent experiments confirms that they are intrinsic to the $5\nu$ oscillation framework rather than artifacts of a particular experimental configuration. The overall sensitivity of T2K is comparable to that of NO$\nu$A, consistent with the two experiments having similar total exposures, with T2K retaining sensitivity down to $\theta_{sa} \approx 0.1$. A notable distinction, however, is that the T2K anti-neutrino channels exhibit appreciably higher sensitivity compared to those of NO$\nu$A, reflecting the relatively more balanced FHC/RHC running of T2K.

\subsection{Combined sensitivity and future prospects}

The current experimental runs of NO$\nu$A and T2K are aiming for substantially higher exposures. NO$\nu$A is presently accumulating data in RHC mode with the goal of approximately doubling the antineutrino exposure by 2027. Similarly, T2K-II targets a total exposure of approximately $10 \times 10^{21}$~POTs, motivated by the goal of achieving $3\sigma$ sensitivity to $\delta_{CP}$.  While the overall topology of the contours remains the same, the improved statistics will sharpen the exclusion boundaries and enhance the significance of features already visible in the current dataset.

\begin{figure}[t!]
    \centering
    \includegraphics[width=0.49\linewidth]{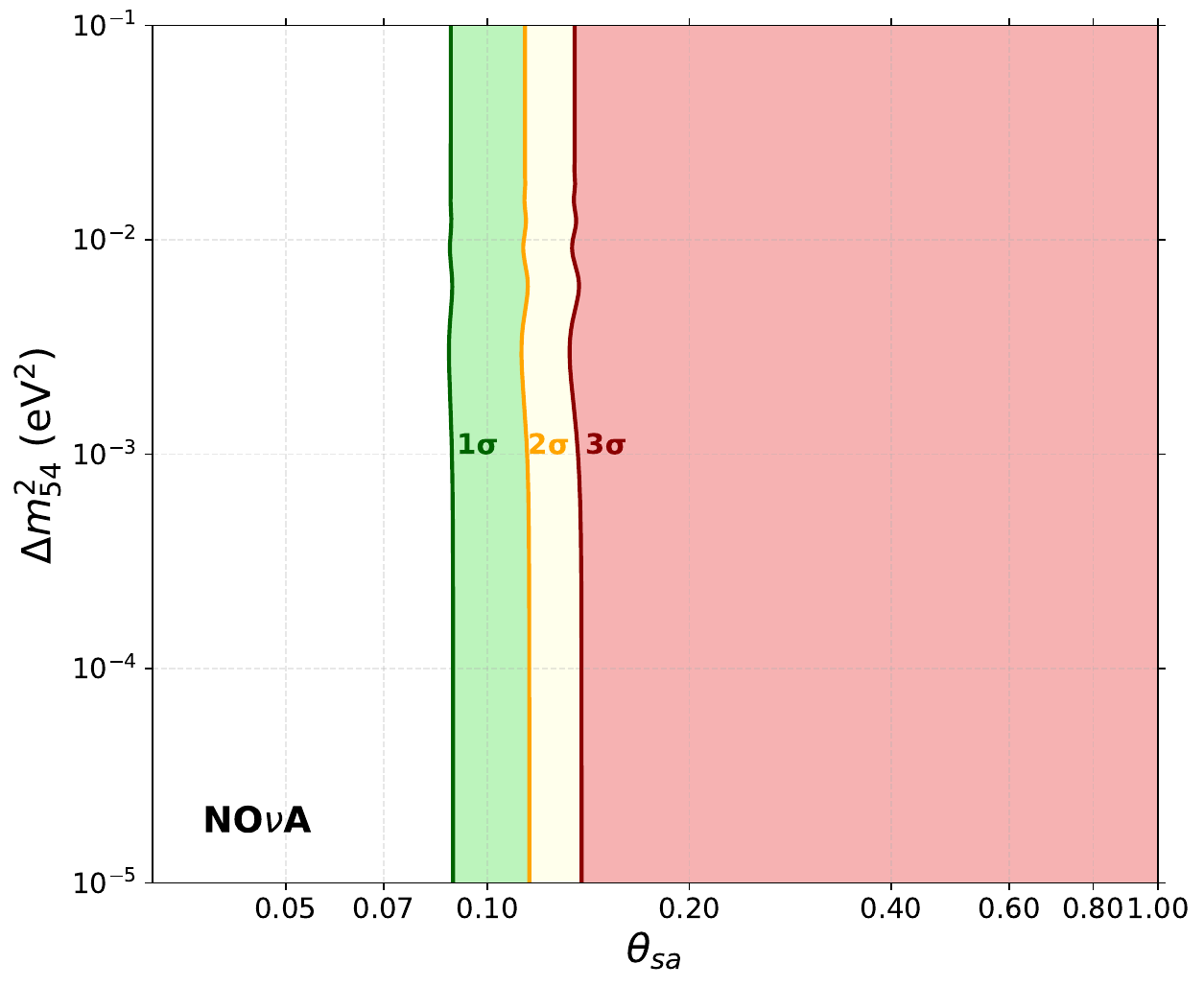}
    \includegraphics[width=0.49\linewidth]{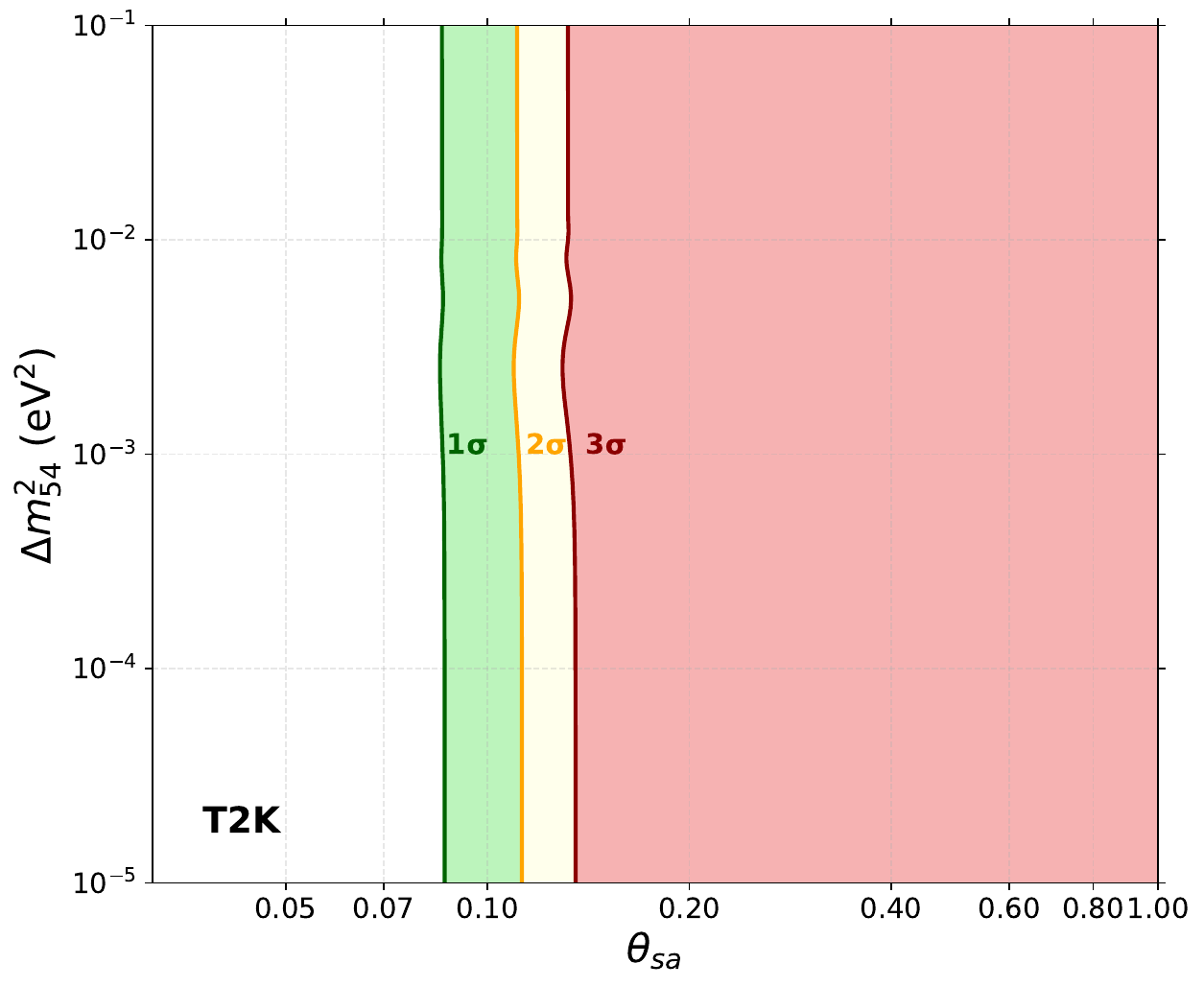}
    \includegraphics[width=0.49\linewidth]{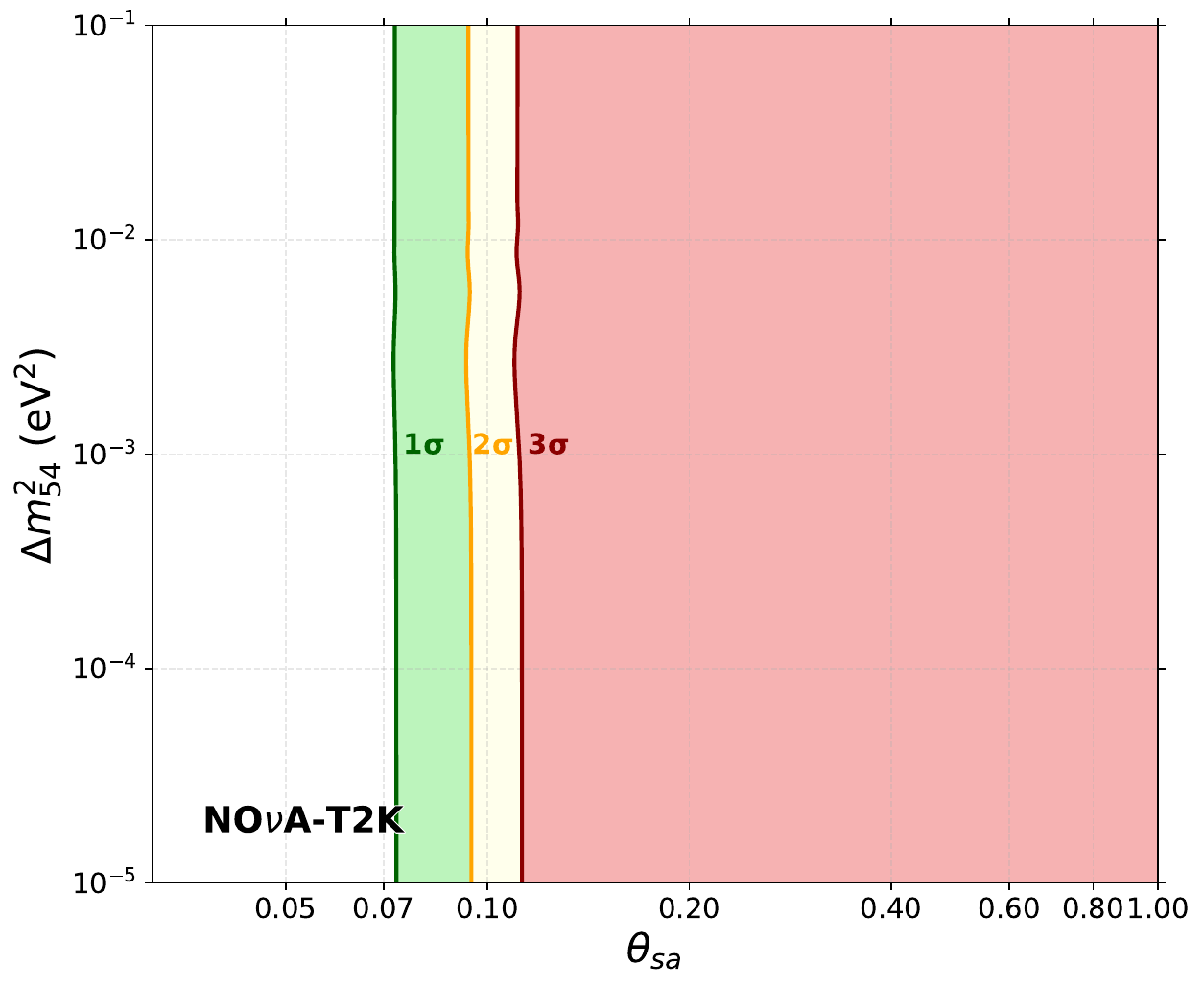}
    \includegraphics[width=0.49\linewidth]{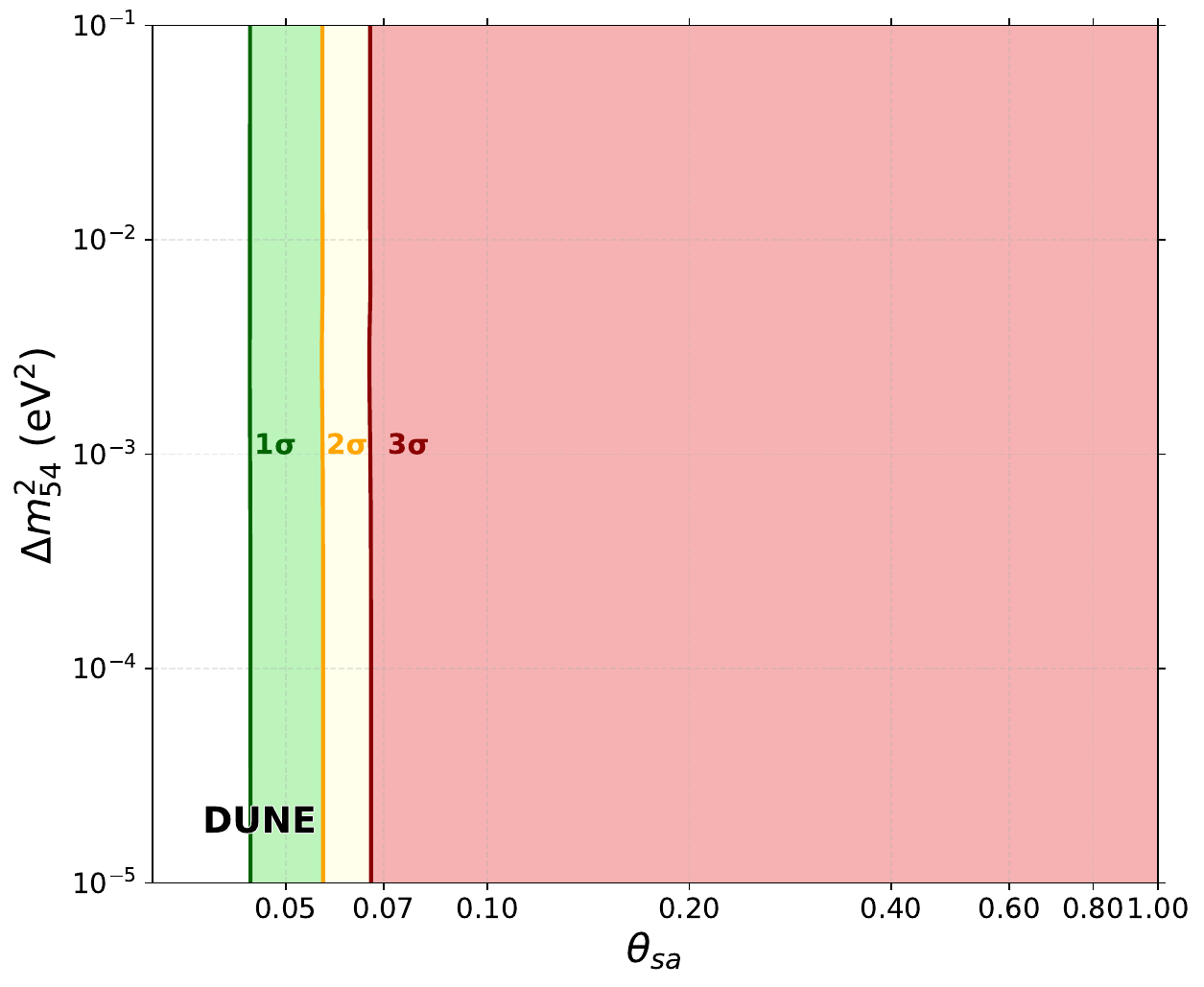}
    \caption{$\chi^2$ sensitivity contour in the plane of mass square splitting $\Delta m^2_{54}$ and sterile mixing $\theta_{st}$ for NO$\nu$A, T2K experiments, Combined NO$\nu$A-T2K and DUNE for all neutrino channels combined. }
    \label{fig:overall_chisq}
\end{figure}

An overall comparison of the combined-channel $\chi^2$ sensitivities is presented in Fig.~\ref{fig:overall_chisq}. A joint NO$\nu$A–T2K analysis yields improved coverage relative to either experiment individually, extending the sensitivity to mixing angles as small as $\theta_{sa} \approx 0.08$. Nevertheless, the sensitivity of DUNE alone exceeds that of the combined NO$\nu$A–T2K analysis across the full mass-splitting range, establishing DUNE as the most powerful prospective probe of quasi-Dirac sterile neutrinos among the experiments considered.
Looking ahead, both NO$\nu$A and T2K are expected to accumulate substantially higher exposures in their ongoing and planned runs. NO$\nu$A is currently operating in RHC mode with a goal of approximately doubling its antineutrino exposure by 2027, while T2K-II targets a total exposure of $\sim 2.0 \times 10^{22}$ POT \cite{T2K:2016siu}, motivated by the goal of achieving $3\sigma$ sensitivity to $\delta_{CP}$. Increasing the NO$\nu$A exposure beyond the current dataset yields only a modest improvement in sensitivity, with the exclusion boundaries tightening marginally across the full range of $\Delta m_{54}^2$ without qualitatively altering the topology of the contours. This suggests that the current NO$\nu$A dataset is already approaching the systematic-limited regime, and that significant further gains in sensitivity to the quasi-Dirac parameter space will require either improved systematic treatment or a next-generation experiment such as DUNE. The sensitivity contours for DUNE are shown in Fig.~\ref{fig:DUNE_4_channels}. In contrast to NO$\nu$A and T2K, the DUNE contours do not exhibit the atmospheric resonance peak near $\Delta m^2_{54} \sim 2.4 \times 10^{-3}~\text{eV}^2$ or the degeneracy valleys near $\theta_{sa} \approx 0.4$. Instead, the sensitivity is approximately uniform across the entire range of $\Delta m^2_{54}$ explored. This behaviour can be attributed to DUNE's significantly longer baseline, over which the rapid sterile-induced oscillations average out across the broad-band energy spectrum. This baseline averaging washes out the resonant interference structures and degeneracy features that are prominent in the near- and mid-baseline experiments. Despite this loss of spectral structure, the uniformity of DUNE's sensitivity represents an advantage: it provides robust, unambiguous coverage of the parameter space without the interpretational complications introduced by degeneracies. Furthermore, DUNE achieves a substantially improved sensitivity to the active-sterile mixing angle, probing down to $\theta_{sa} \approx 0.05$, roughly a factor of two improvement over NO$\nu$A and T2K.

\subsection{CP-violation dependence and the five-neutrino framework} 
\label{sec:deltacp_5nu}
As discussed in Section~\ref{QD_LBL_Theory}, the CP-violating phase $\delta_{CP}$ plays a significant role in shaping the $\chi^2$ sensitivity. At $\delta_{CP} = 0.82$, the CP-dependent term in the $3\nu$ appearance probability is destructive, and the $5\nu$ model, by suppressing this contribution through the factor $c_{sa}$, drives its prediction closer to a CP-conserving baseline. The spectral difference between the $3\nu$ and $5\nu$ models is therefore reduced, and with it, the discriminating power of the $\chi^2$ test.

\begin{figure}[t!]
    \centering
    \includegraphics[width=0.9\linewidth]{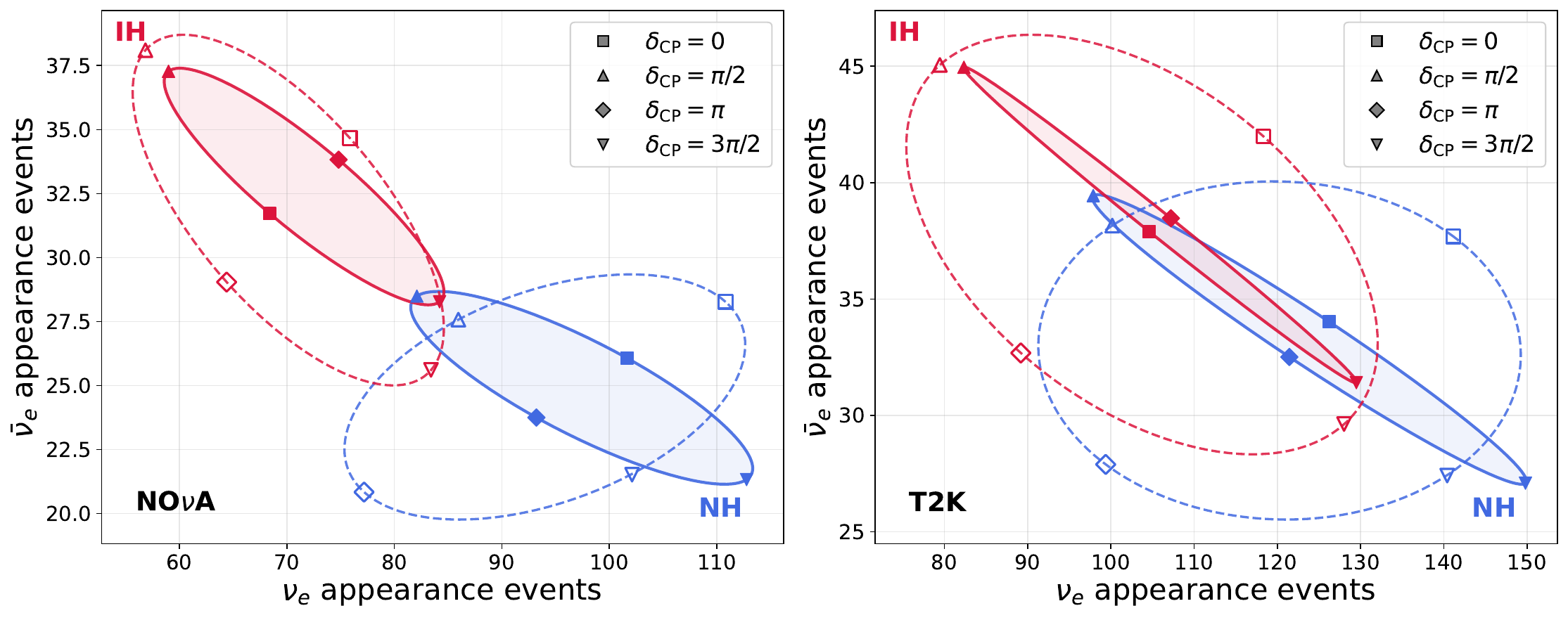}
    \caption{Comparison of the predicted bi-event rates for the $3\nu$ (solid) and $5\nu$ (dashed) frameworks at NO$\nu$A and T2K for both normal and inverted mass orderings. The four shapes correspond to the respective $\delta_{CP}$ values $(0,\frac{\pi}{2},\pi,\frac{3\pi}{2})$ for $\Delta m_{41}^2=\text{10 eV}^2$, $\Delta m_{54}^2=10^{-2}\text{ eV}^2$ \text{ and } $\theta_{sa} = 0.1$.}
    \label{fig:nova_t2k_nhih}
\end{figure}
To explore the interplay between $\delta_{CP}$ and the quasi-Dirac sterile sector, we use a bievent representation of the predicted event rates. Similar to the oscillation probability, each integrated event rate can be decomposed into contributions proportional to $\sin\delta_{CP}$ and $\cos\delta_{CP}$ as shown in \ref{eq:CP_Pro_Expression}. Since $\delta_{CP}$ remains one of the least constrained parameters in the standard 3$\nu$ framework, it can be varied continuously over $[0,\, 2\pi]$. The locus of predicted points $(N_{\nu_e},\, N_{\bar{\nu}_e})$ traces out an ellipse in the bievent plane as a direct consequence of the sinusoidal dependence on a single angular parameter. Every point along it corresponds to a physically allowed prediction for some value of $\delta_{CP}$. In the $5\nu$ framework, an analogous ellipse can be constructed for each choice of the sterile parameters $(\theta_{sa},\, \Delta m^2_{54})$. 

The comparison across both normal and inverted mass orderings is shown in Fig.~\ref{fig:nova_t2k_nhih}. The $5\nu$ data points lie within the experimental uncertainty bands of the $3\nu$ expectations. However, this apparent consistency should be interpreted with caution. While the $5\nu$ predictions individually overlap with the $3\nu$ uncertainty envelope, the overlap among different $5\nu$ parameter choices is substantially larger than the spread among distinct $3\nu$ configurations. In other words, a wider range of sterile parameter combinations can produce event spectra that are mutually compatible, introducing additional degeneracies that complicate parameter extraction.

\section{Conclusion}

We have presented a comprehensive analysis of quasi-Dirac neutrino oscillations at long-baseline experiments within an extended five-neutrino framework motivated by the inverse seesaw mechanism. Extending the Standard Model with two SM-singlet Weyl fermions, a small lepton-number-violating Majorana mass $\mu$ generates a quasi-Dirac sterile pair with mass splitting $\Delta m^2_{54}$, leaving the active sector well described by the standard PMNS framework. We derived analytic oscillation probabilities for the $\nu_\mu \rightarrow \nu_e$ and $\nu_\mu \rightarrow \nu_\mu$ channels including MSW matter effects, and map sensitivity across the $(\theta_{sa}, \Delta m^2_{54})$ parameter space using \texttt{GLoBES} simulations of NO$\nu$A, T2K, and DUNE with full systematic treatment.

Several structural features of the $5\nu$ oscillation framework were identified consistently across experiments. In the degenerate limit $\Delta m^2_{54} \rightarrow 0$, the sterile pair mimics a single Dirac neutrino and sensitivity vanishes. At intermediate splittings, a resonance enhancement near $\Delta m^2_{54} \sim \Delta m^2_{31} \sim 2.4 \times 10^{-3}$ eV$^2$ produces pronounced spectral distortions at NO$\nu$A and T2K, but simultaneously induces parameter degeneracies with the atmospheric sector. A second degeneracy near $\theta_{sa} \approx \theta_{23} \approx 0.4$ allows the $5\nu$ model to reproduce the $3\nu$ spectral shape, further limiting discriminating power. Comparison of $3\nu$ and $5\nu$ event rate predictions confirmed that current NO$\nu$A and T2K data cannot exclude the quasi-Dirac scenario across the full range of benchmark CP phases and mass orderings, with sensitivity limited to $\theta_{sa} \gtrsim 0.1$ individually and $\theta_{sa} \gtrsim 0.08$ in combination.

The sensitivity of DUNE represents a qualitative improvement. Its 1300 km baseline, 40 kt fiducial mass, and broad-band beam yield uniform coverage down to $\theta_{sa} \approx 0.05$ across the entire $\Delta m^2_{54}$ range, free from the resonance structures and degeneracy valleys that characterise the current generation. This robustness follows from the averaging of rapid sterile-induced oscillations over the broad energy spectrum, with sensitivity driven primarily by the overall suppression of active neutrino fluxes. We further demonstrated that the quasi-Dirac sensitivity is intrinsically coupled to the measurement of $\delta_{CP}$: the $5\nu$ appearance probability contains a suppression factor $c_{sa}$ modulating the CP-dependent amplitude, such that the discriminating power depends sensitively on whether CP-odd interference is constructive or destructive in the standard three-flavour case. This mutual dependence implies that DUNE's quasi-Dirac and CP-violation programs will necessarily inform one another.

Several important directions for future work remain. Relaxing the flavour-universality assumption on sterile-active mixing would substantially enrich the phenomenology, enabling independent searches across the $\nu_e$, $\nu_\mu$, and $\nu_\tau$ appearance channels. The impact of non-zero sterile CP phases, set to zero throughout this analysis, warrants systematic investigation, as these could either resolve or deepen the degeneracies identified here. Complementary constraints from neutrinoless double beta decay, where quasi-Dirac pairs produce near-complete cancellations in the effective Majorana mass, with observable residuals accessible at nEXO and LEGEND, and from high-energy astrophysical neutrinos at IceCube, sensitive to mass splittings far below the reach of terrestrial baselines, would provide powerful independent probes within the same framework.
This study establishes that long-baseline oscillation experiments offer meaningful and complementary sensitivity to quasi-Dirac neutrino oscillations. DUNE, with its superior baseline and exposure, will provide the definitive near-term probe of this scenario, pushing sensitivity to $\theta_{sa} \approx 0.05$ with unambiguous parameter space coverage. As the global programme advances through T2K-II, the extended NO$\nu$A run, and the full DUNE dataset, the combination of these experiments will place increasingly stringent constraints on lepton number violation in the neutrino sector, directly bearing on the fundamental question of the Dirac or Majorana nature of neutrino mass.

\acknowledgments
F.~F.~D. and N.-I.~B. acknowledge support from the UK Science and Technology Facilities Council (STFC) via the Consolidated Grant ST/X000613. N.-I.~B. would like to thank the Wright Laboratory at Yale University, where part of the this work was conducted, for their hospitality.

\bibliography{neutrinos}
\end{document}